# Ab initio quantum many-body description of superconducting trends in the cuprates

Zhi-Hao Cui,[1,*] Junjie Yang,[1] Johannes Tölle,[1] Hong-Zhou Ye,[2] Shunyue Yuan,[3] Huanchen Zhai,[1] Gunhee Park,[3] Raehyun Kim,[4] Xing Zhang,[1] Lin Lin,[4,5] Timothy C. Berkelbach,[2] and Garnet Kin-Lic Chan[1,†]

[1]*Division of Chemistry and Chemical Engineering, California Institute of Technology, Pasadena, California 91125, USA*
[2]*Department of Chemistry, Columbia University, New York, New York 10027, USA*
[3]*Division of Engineering and Applied Science, California Institute of Technology, Pasadena, California 91125, USA*
[4]*Department of Mathematics, University of California, Berkeley, California 94720, USA*
[5]*Computational Research Division, Lawrence Berkeley National Laboratory, Berkeley, California 94720, USA*

Using a systematic ab initio quantum many-body approach that goes beyond low-energy models, we directly compute the superconducting pairing order and estimate the pairing gap of several doped cuprate materials and structures within a purely electronic picture. We find that we can correctly capture two well-known trends: the pressure effect, where the pairing order and gap increase with intra-layer pressure, and the layer effect, where the pairing order and gap vary with the number of copper-oxygen layers. From these calculations, we observe that the strength of superexchange and the covalency at optimal doping are the best descriptors for these trends. Our microscopic analysis further identifies that strong short-range spin fluctuations and multi-orbital charge fluctuations drive the development of the pairing order. Our work illustrates the possibility of a material-specific ab initio understanding of unconventional high-temperature superconducting materials.

## I. INTRODUCTION

Since the discovery of high-temperature superconductivity in the cuprates almost 40 years ago, obtaining a microscopic description of the phenomenon has challenged theoretical material science [1, 2]. In particular, the search for new materials with higher transition temperatures has been hindered by the absence of predictive computational links between the material structure/composition and the observed superconducting temperatures. Here, we describe microscopic calculations that reproduce some of the best-known material trends in cuprate superconducting critical temperatures $T_c$ via the direct ab initio computation of the ground-state pairing order, using only the material structure as input. These rely on new methods to solve the quantum many-body Schrödinger equation in the materials without first simplifying to low-energy models. Analyzing the solutions identifies simple descriptors which correlate with transition temperature and the fluctuations that drive the microscopic process of pairing. Overall, our methodology demonstrates a path towards predictive ab initio computations of high-temperature superconductivity in new materials.

Cuprate superconductors are layered perovskite compounds with two-dimensional copper-oxygen planes separated by buffer layers of atoms, which dope the planes either with electrons or holes. In the parent undoped state, the materials are antiferromagnets, becoming superconducting after doping beyond $\sim$5% [3, 4]. Out of the many efforts to increase $T_c$ through altering the composition and structural parameters, some trends can be identified. Two of the clearest ones are the pressure effect and layer effect. In the pressure effect, onset $T_c$ increases with pressure applied in the plane, rising, e.g. in Hg-1223 from 135 K at ambient pressure to 164 K at 30 GPa [5, 6]. In the layer effect, $T_c$ increases with the number of stacked copper-oxygen planes (e.g. in the mercury-barium cuprates, $T_c$ is 97, 127, 133 K in the 1-, 2-, 3-layer compounds [7]).

Many theories have been proposed to rationalize cuprate superconductivity, but it has proven difficult to obtain a detailed microscopic picture, and even harder to reproduce the specifics of different cuprate materials. There are two essential complications. First, the phenomenon arises from quantum many-body physics with strong electron interactions, where there are no analytical solutions and there is no obvious small parameter [8]. (This is in contrast to conventional superconductors with weak electron interactions, where material-specific computations are relatively successful [9, 10]). Within any microscopic framework to describe the electron correlation, the predictions thus carry uncertainty from their approximate nature. The second is that the complex material composition complicates the derivation of low-energy Hamiltonians. While one-band Hubbard models and their relatives have informed much current thinking [11, 12], recent accurate numerical solutions of these models have also highlighted the deviation of the model physics from that of the real materials and the sensitive dependence of the physics on the model [13–18]. In addition, although there has been progress in rationalizing material specific effects in terms of parametrized multi-band models [12, 19–21] the uncertainty introduced into the Hamiltonian arising from downfolding, for instance, due to density functional theory double-counting [22], the definition of impurity orbitals [23], or the difficulties of parametrization or uncertainty of the parametrized form [24], appears comparable to the strength of the material trends.

In principle, solving the ab initio many-electron Schrödinger equation for the full cuprate material provides an unambiguous and material-specific route to understanding cuprate superconductivity. Although this is traditionally viewed as intractable, recent advances in numerical many-body algorithms and their computational implementation are opening up the possibility of predictive ab initio computation even in strongly correlated quantum materials. While such calculations are more expensive than their model counterparts and thus more limited in the system size that can be treated, they are complementary to the low-energy model approach as the microscopic Hamiltonian unambiguously reflects the material-specific composition, at the expense of a less detailed description of long-range physics. As one example of the success of such a strategy, we previously captured, and illuminated at the atomic level, systematic trends in the magnetism of the parent state of the cuprates with such an approach [25]. Here, we show how these strategies may be extended to the much more challenging doped phases of the cuprates, and in particular to obtain the superconducting pairing order and an estimate of the pairing gap. Below, we describe the advances that now make this work possible, the cuprate systems we will study for their systematic trends, and the results and insights that derive from this approach.

* zhcui0408@gmail.com; Present address: Department of Chemistry, Columbia University, New York, New York 10027, USA
† gkc1000@gmail.com

## II. QUANTUM SIMULATION METHODS

We aim to approximate, ab initio, the ground-state of the electronic Schrödinger equation of the bulk cuprate (phonon and temperature effects are thus ignored). The strategy has three pieces: a quantum embedding (density matrix embedding theory (DMET)) to connect the bulk many-body problem to a self-consistent impurity many-body problem; the quantum chemical solution of the impurity problem; and the quantum chemical mean-field solution of an auxiliary bulk problem. To retain material specificity, the Hamiltonians use ab initio bare electronic interactions, expressed in a basis of a few hundred of bands, thus no reduced models appear.

Density matrix embedding theory has been introduced elsewhere [26], and its application to doped Hubbard models [13, 27, 28] and ab initio cuprate parent states, extensively benchmarked [25]. We briefly recount essential details in Fig. 1. DMET provides a zero-temperature quantum embedding that maps the interacting bulk problem to the self-consistent solution of two systems: an interacting quantum impurity and an auxiliary mean-field bulk problem. The quantum impurity is taken as a computational supercell (with all atoms) of the material, coupled to a bath constituting the most important orbitals of its environment. The auxiliary bulk Hamiltonian $H^{\text{latt}}$ is a mean-field crystal Hamiltonian, augmented by a one-electron operator $\Delta$ in each unit cell. The mean-field ground-state $\Phi^{\text{latt}}$ (a Slater determinant or Bardeen-Cooper-Schrieffer (BCS) state, depending on $\Delta$) determines the bath orbitals, and thus the embedding (impurity plus bath) Hamiltonian $H^{\text{emb}}(\Delta)$. The impurity ground-state $\Psi^{\text{emb}}$ then determines the order parameter $\kappa^{\text{emb}}$. The quantum impurity and auxiliary bulk problems are solved self-consistently with respect to $\Delta$ until $\kappa^{\text{latt}}(\Delta) = \kappa^{\text{emb}}(\Delta)$. $\Delta$ and $\kappa$ can acquire finite values for ordered phases due to symmetry breaking in the self-consistency.

In an ab initio description, we start with an atomic orbital representation of the crystal. For bases with reasonable accuracy, this gives rise to many bands, e.g. a few hundred bands per computational cell. The quantum impurity, which contains the orbitals of the atoms in the impurity, thus also contains hundreds of orbitals, and we require an ab initio many-body treatment for such problems. Fortunately, only a few orbitals are strongly correlated, so we can use quantum chemistry strategies designed to handle hundreds of orbitals with a few strongly correlated ones: here we primarily use the coupled cluster singles and doubles (CCSD) approximation [29]. Such coupled cluster wavefunctions are widely used in molecular, and more recently materials, modeling and have proved accurate for ordered states (in the current setting we additionally verify their accuracy through other quantum chemical methods, such as the ab initio density matrix renormalization group [30]).

In the context of the doped phases of the cuprates, new ingredients appear, such as the treatment of doping. In real materials, doping usually involves dopant atoms, which enlarge the computational cell [31]. For simplicity, we use implicit doping which modifies the charge density while introducing a compensating positive field, either within the "rigid-band" approximation (RBA, uniform background field), or (for a subset of calculations), the virtual crystal approximation [32] (VCA, scaled external field at select atoms). These treatments of doping are undoubtedly crude. Although detailed results for individual structures are sensitive to the doping formulation, we find trends across the materials to be preserved within a fixed doping scheme.

Another new ingredient is the ab initio simulation of superconducting phases. Defining $\Delta = \sum_{ij} \Delta_{ij} a_i^\dagger a_j^\dagger + \text{H.c.}$ and $\kappa_{ij} = \left\langle a_j^\dagger a_i^\dagger \right\rangle$, past a critical doping, the DMET self-consistency produces a finite $\Delta$ and $\kappa$. Such superconducting solutions are not usually supported by ab initio quantum chemistry solvers. To handle this, we use the Nambu-Gorkov formalism [33] which, for $S_z = 0$ pairing, maps broken particle number symmetry to broken $S_z$ symmetry. This creates a particle-conserving $H^{\text{emb}}$ amenable to standard quantum chemistry methods, as further detailed in Sec. 1 of [34].

The scale of the simulations in this work also required additional innovations. For example, to achieve an affordable description of the quantum many-body state, we developed new, compact, Gaussian atomic bases, of correlation-consistent double-$\zeta$ plus polarization quality. Similarly, to treat doped states which may be metallic, we adapted our orbital localization, self-consistency procedures, and solver algorithms for metallic systems. These and other technical improvements are discussed in Sec. 1 of [34].

The outputs of the ab initio DMET procedure are a correlated quantum impurity wavefunction $\Psi^{\text{emb}}$ and a mean-field bulk wavefunction $\Phi^{\text{latt}}$. The former can be used to obtain impurity observables, such as the pairing order, while the latter provides additional information (albeit of limited fidelity) on long-range non-local observables. $\kappa$ is a multi-orbital quantity, and can be summed into a scalar order parameter with different angular symmetry (in various ways due to the multi-orbital character, see Eqs. (S71) and (S72)); we use $m_{\text{SC}}$ to denote the scalar summed quantities. The impurity fluctuations that give rise to these orders can be analyzed from $\Psi^{\text{emb}}$.

Because the pairing order is not easily measurable, we also consider additional quantities to place our results in an experimental context. We compute the maximal pairing gap $E_{\text{g}}$ (approximately corresponding to $\sim 2\Delta_0$ in a 1-band BCS theory) of the auxiliary bulk Hamiltonian $H^{\text{latt}} + \Delta$. We note that this is not the same as computing the pairing gap of the interacting bulk problem, but we are using it to provide information in a manner analogous to how a DFT bandgap is used to provide information on the true bandgap of the problem. In the weak-coupling regime (which may not include the cuprates) $2\Delta_0$ is also proportional to $T_c$ (see Sec. 3.7 of [34]).

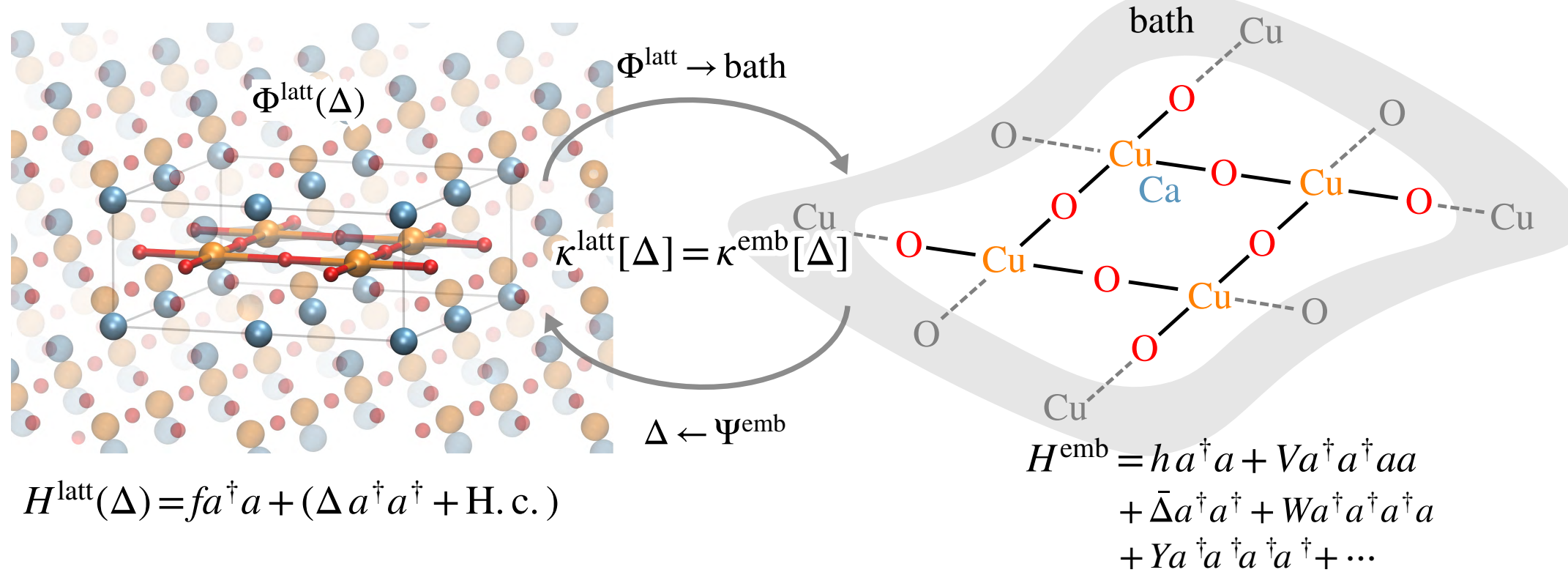

FIG. 1. **Computational strategy.** The ab initio density matrix embedding theory (DMET) framework. This involves solving two ground-state problems: for an auxiliary mean-field Hamiltonian (left), $H^{\mathrm{latt}} = f + \Delta \to \Phi(\Delta)$, and a quantum impurity + bath Hamiltonian (right), $H^{\mathrm{emb}}(\Delta) \to \Psi^{\mathrm{emb}}(\Delta)$. $\Delta$ is modified by self-consistent iteration until the pairing order $\kappa$ is the same in the impurity and the auxiliary mean-field problem. The non-number-conserving $\bar{\Delta}, W, Y$ terms in $H^{\mathrm{emb}}$ arise from the DMET bath construction from $\Phi(\Delta)$. In this work, the bulk problem is represented by 128 cuprate unit cells, and the impurity is a $2 \times 2$ supercell, illustrated above for CCO ($CaCuO_2$).

Within the above formulation, the errors can be attributed to the following sources: the finite impurity supercell, the finite atomic orbital basis, approximations in the many-body solver, and approximations in the DMET self-consistency. In principle, these errors can be improved to exactness, as has been analyzed in [25]. In practice, the errors remain finite in our computation. For example, the finite impurity supercell in this work means that we omit some interesting long-range physics, such as stripes or other long-wavelength orders. Our calculations may then be viewed as asking, if we restrict ourselves only to orders that can be formed in the finite cell, is the physics sufficient to explain material-specific trends? Below, we will examine such questions through computation.

## III. CUPRATE SYSTEMS AND COMPUTATIONS

We consider two series of hole-doped cuprates to study the pressure and layer effects. The first is CCO ($CaCuO_2$), viewed as a parent compound for a variety of doped cuprates. When mixed with Sr, it has been doped with vacancies (approximate composition $(Ca_{1-y}Sr_y)_{1\text{-}\delta}CuO_2$ with $y \sim 0.7$, $\delta \sim 0.1$, $T_c \sim 110$ K [35]). The simple structure of parent CCO makes it ideal for studying the pressure effect. We apply pressure along the $a, b$ axes of the cuprate plane in a manner which can be compared to the uniaxial pressure derivatives $\mathrm{d}T_c/\mathrm{d}P_a$, $\mathrm{d}T_c/\mathrm{d}P_b$ extracted in [36]. (We note that in the hydrostatic experiments of Ref. [6], the trend of increasing $T_c$ with pressure (positive $\mathrm{d}T_c/\mathrm{d}P$) was observed for the onset $T_c$ across a series of mercury cuprates [5], but for $T_c$ corresponding to bulk zero resistance, was found to be compound specific [6]. Given that the small size of our simulation cells does not allow for a full treatment of phase fluctuations, capturing the difference between onset and zero resistance $T_c$ is beyond our current description).

The second series of compounds are the mercury barium cuprates, single-layer Hg-1201 ($HgBa_2CuO_{4+\delta}$) and double-layer Hg-1212 ($HgBa_2CaCu_2O_{6+\delta}$). In the experimental setting, this series can be continued up to 6-layers, but the computational cell of these systems is too large for a practical treatment. Consequently, to give some information on the large layer limit, we consider CCO as a crude comparison to form the infinite layer parent compound in this series. The mercury-barium compounds are synthesized under conditions with finite oxygen doping. In Hg-1201 we use a structure corresponding to a reported oxygen doping $\delta = 0.19$, associated $T_c \sim 95$ K [37]; in Hg-1212, we consider two structures [38]: the oxidized structure ("ox", with $\delta = 0.22$, $T_c \sim 128$ K), and a variant argon-reduced structure ("red", close to the undoped compound, $\delta = 0.08$, $T_c \sim 92$ K).

The computations start with a mean-field density functional (DFT) calculation using the Perdew-Burke-Ernzerhof with exact exchange (PBE0) functional in a custom polarized double-zeta Gaussian basis (described in Sec. 2.2 of [34]) sampling the Brillouin zone corresponding to $8 \times 8 \times 2$ **k**-points of the primitive cell. (More precisely, we use supercells, thus using a $2 \times 2$ supercell, we sample the corresponding $4 \times 4 \times 2$ folded Brillouin zone). The increment of doping in the DMET calculation derives from the size of the bulk calculation: we can dope in units of $1/128$. Because we do not perform full self-consistency on the charge density (see below) it is necessary to use a mean-field starting point that produces a reasonable charge distribution. We select the PBE0 functional based on its performance relative to accurate quantum many-body embedding benchmarks in our previous work on the parent state [25]. The PBE0 solution is antiferromagnetic (AFM) at half-filling, and in some cases, becomes paramagnetic beyond a certain doping. The PBE0 calculation generates the initial auxiliary mean-field Hamiltonian $f$ (Fig. 1), but this does not enter the many-body impurity Hamiltonian $H^{\mathrm{emb}}$. Thus the correlated DMET calculations do not have a double counting error.

The quantum impurity problem consists of the $2 \times 2$ cell of the cuprate and the DMET bath constructed for the valence orbitals in the impurity, yielding a total problem size of 300-900 orbitals. The upper range corresponds to the large cells of the mercury-barium compounds, and to reduce the cost we used coupled sub-impurities to separately treat the $CuO_2$ and buffer layers [25]; the largest sub-impurities contain 376 orbitals. As already discussed above, the $2 \times 2$ cell means that we omit long wavelength physics, such as stripes, which we discuss further in [34] Sec. 4. For the results discussed below, we show data from CCSD as a compromise between speed and accuracy. Benchmarks of CCSD against exact solvers in the parent state [25], in the DMET treatment of the 2D one-band Hubbard model and in the ab initio cuprate impurity problem (Figs. S3 and S4) suggest we reach sufficient accuracy to discuss the material trends of interest.

To simplify the convergence of the self-consistency, we carry it out with respect to the pairing potential $\Delta$ restricted to the three-band Cu $3d_{x^2-y^2}$ and O $2p_{x(y)}$ orbitals with matrix elements restricted to obey $C_{2h}$ point group symmetry. This allows for the direct update of the pairing density $\kappa$ (to self-consistency), although it limits self-consistency on the normal charge density itself. In addition, we do not update the charge contribution to the mean-field $f$. Without full charge self-consistency, the converged DMET solution retains a dependence on the initial choice of mean-field $f$ and initial density. The effect of this dependence is discussed more below and in Sec. 3.8 of [34].

## IV. THE PRESSURE EFFECT

We first examine the computed order in CCO at three different in-plane pressures: -19, 0 (ambient), 32 GPa. Without doping, CCO is in an antiferromagnet. At all pressures, under sufficient doping, a superconducting state is formed with predominantly $d$-wave pairing order, as illustrated by the Cu-Cu pairing order at optimal doping in Fig. 2B; the $d$-wave character increases with increased pressure. There is also a small $s$-wave piece in the Cu-Cu pairing, and the total order has a small $p$-wave component necessarily arising from the coexistence of AFM and $d$-wave superconducting order [16, 39]. The uncertainties of the calculation mean that the absolute numerical values for the order should be treated with caution. However, we can obtain a conversion to experimentally accessible quantities through the computed pairing gap amplitude $E_{\mathrm{g}}$ (right-hand axis of Fig. 2E). The pairing gap closely follows the pairing order and the maximum shows the same trend with pressure. Using the weak-coupling result (with its associated limitations) for the conversion of the gap to $T_{\mathrm{c}}$ (Sec. 3.7 of [34]), we obtain $T_{\mathrm{c}} \sim 180$ K at ambient pressure, of the same order as that typically seen in experiment. Then $\mathrm{d}T_{\mathrm{c}}/\mathrm{d}P \approx 4-5$ K/GPa, comparable to the value extracted from the uniaxial pressure derivatives $\mathrm{d}T_{\mathrm{c}}/\mathrm{d}P \sim 2\mathrm{d}T_{\mathrm{c}}/\mathrm{d}P_a \sim 4-6$ K/GPa [36]. Our calculations thus capture the qualitative structural trend of pressure on the maximum superconducting order and temperature.

Fig. 2C shows a real-space visualization of the Cu-centered pair amplitude, and the scalar pair amplitude between orbitals on neighboring Cu atoms, for CCO at ambient pressure and optimal doping. The sign of the pairing amplitude illustrates the $d$-wave symmetry, while the spread (not shown) corresponds to a pair distributed across a linear distance of about 6 unit cells. We show a more detailed orbital resolved analysis of the Cu-Cu $d$-wave order at optimal doping in Fig. 2D. As illustrated in Fig. S5, as doping increases the pairing orbital character changes, with Cu-Cu pairing at small doping being predominantly $3d$-$3d$, but at larger dopings containing more $4s$ and $4d$ components. We find that O-O pairing contributes about 30% to the total $d$-wave order, with Cu-O pairing contributing mainly to the $p$-wave order.

We now examine in detail the AFM and SC orders as a function of doping in Fig. 2E. We first discuss the $x$-axis, the doping axis. When holes are added, the charges go primarily to the $CuO_2$ plane, and reside mainly on oxygen ($2p$), with a fraction (about 20% - 30%) transferred to Cu; about 90% of the charge resides in the three-band orbitals (Cu $3d_{x^2-y^2}$ and O $2p_{x,y}$), and about 97% in the Cu-O plane. In experiments on oxygen doped cuprates, such as yttrium barium copper oxide [40] and the mercury barium cuprates studied later [41], the effective Cu doping is usually not taken from the estimated oxygen content, which (depending on the assumed formal charge of the dopants) could translate to very large copper-oxygen plane dopings. Instead, the effective doping of the copper-oxygen plane is inferred from an empirical formula [40, 41]. This empirical formula suggests that the optimal doping of the plane is usually about 10%-15% by which the magnetic order is seen to vanish.

In contrast, we see that the local moment in our calculations decays quite slowly and the SC order appears only at much larger dopings. This primarily reflects the residual dependence of our non-charge-self-consistent DMET calculations on the initial DFT charge density, which also has a similarly slow decay of the local moment. (An extreme example of this is -19 GPa system, where the moment does not decrease even under heavy doping).

While full charge self-consistency was too expensive to explore here, we can understand its potential corrective effect in a 3-band model (see Sec. 3.8 of [34]). When doping holes into the model, they go into a pair of (i.e. $\uparrow$ and $\downarrow$) spin-polarized bands that have mixed Cu-O character. Thus some of the $\uparrow$ and $\downarrow$ spin-hole density overlaps on the oxygen atoms (with compensated moments) and the residual cell moment comes only from the uncompensated moment on Cu. As shown in Sec. 3.8 of [34], allowing for charge self-consistency changes the Cu-O character in the 3-band model. This leads to a much faster decay of the magnetic moment.

It is interesting to replace the bare doping here with the hole content on Cu computed from the atomic populations, $x_{\mathrm{Cu}}$. Using this rescaling, we empirically observe that the maximum $d$-wave order appears at a Cu hole content (10%-15%) similar to that seen in one-band treatments [42], as well as in experiment. The optimal doping is close to the point at which the magnetic order suddenly drops.

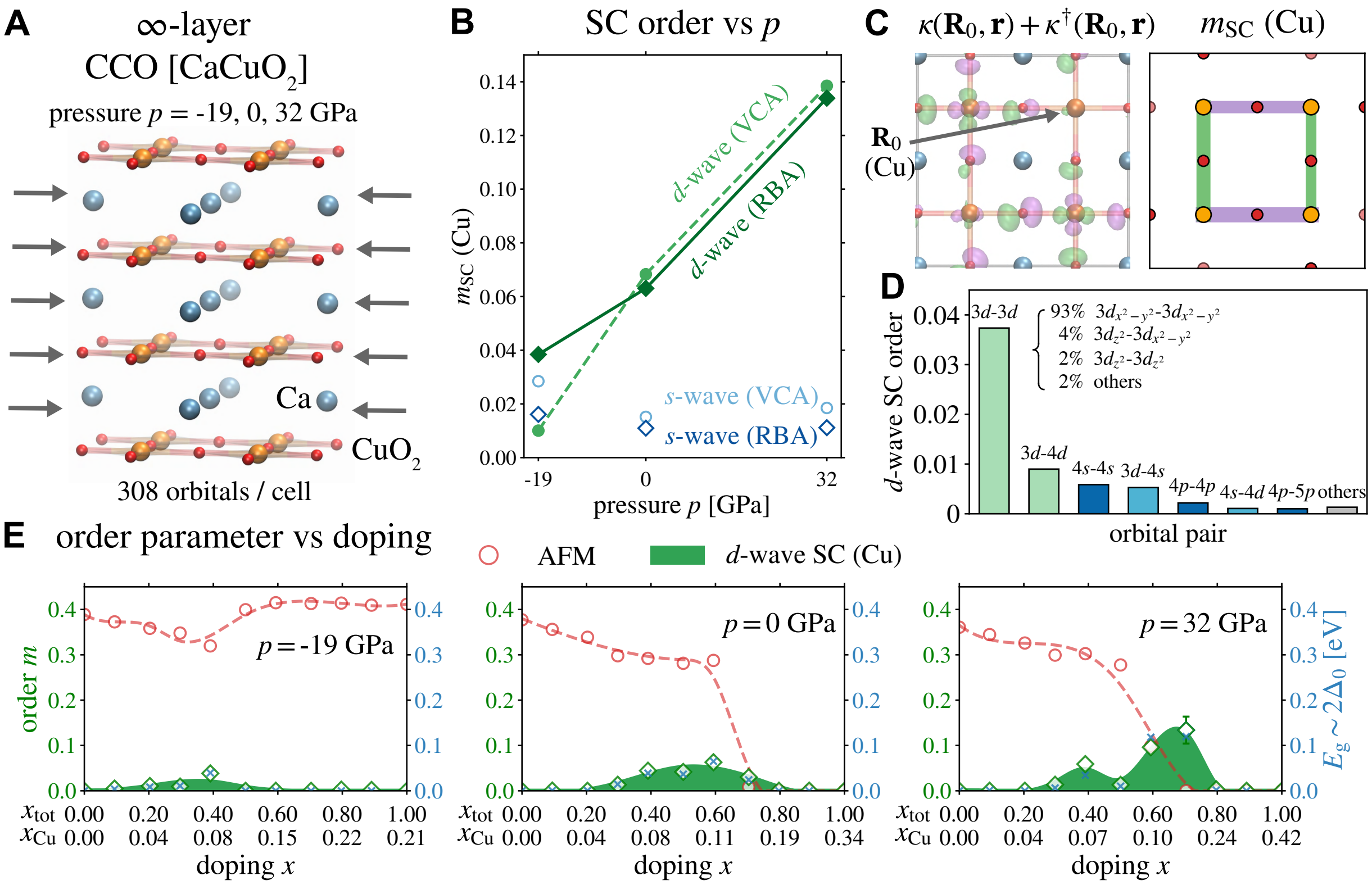

FIG. 2. **Superconducting order and pressure effect.** (A) Structure of the $\infty$-layer cuprate CCO ($CaCuO_2$). (B) $d$-wave and $s$-wave order as a function of pressure $p$, using different doping representations (rigid band approximation (RBA) and virtual crystal approximation (VCA)). (C) Anomalous density $\kappa(\mathbf{R_0}, \mathbf{r}) + \kappa^{\dagger}(\mathbf{R_0}, \mathbf{r})$ for CCO at optimal doping and ambient pressure. The reference point $\mathbf{R_0}$ is near the Cu atom in the embedded cell. $m_{\mathrm{SC}}$ (Cu): pairing order between neighboring Cu atoms showing $d$-wave symmetry. (D) Orbital-resolved $d$-wave SC orders between Cu orbital pairs. (E) AFM, SC order $m$ and estimated SC gap $E_{\mathrm{g}}$ as a function of doping using RBA (VCA curves shown in Fig. S6). The error bar at $x = 0.7$ doping at 32 GPa is also shown due to the slow convergence of DMET.

The numerical values of the magnetic and pairing orders depend on details of the doping treatment: for example, the difference in the maximum pairing order between a VCA and RBA treatment is shown in Fig. 2B and Fig. S6. This highlights the need to investigate more realistic representations of dopants. However, the qualitative pressure trend in the pairing order is reproduced in either case.

## V. THE LAYER EFFECT

We next consider the layer effect in the mercury barium cuprates (1-, 2-layer) using CCO as a proxy for the $\infty$-layer compound. The plot of the maximum pairing order as a function of the layer number is shown in Fig. 3C. We see a sizable increase in the maximum pairing order moving from Hg-1201 to Hg-1212 (ox). The maximum pairing gap $E_{\mathrm{g}}$ (right axis of Fig. 3D) shows the same behavior as the pairing order. Both are similar to the experimental change in $T_{\mathrm{c}}$. Again, although the proportionality between $E_{\mathrm{g}}$ and $T_{\mathrm{c}}$ is a weak-coupling relation, arguably we seem to capture the basic experimental trend in $T_{\mathrm{c}}$ of the layer effect. For more than 3 layers, experimentally it is seen that $T_{\mathrm{c}}$ no longer increases, which has been attributed to the potentially inhomogeneous doping of the different copper-oxygen planes. In CCO, inhomogeneous doping is not part of our representation. However, assuming CCO is a reasonable analog for the $\infty$-layer limit of the mercury barium cuprates, we find that the pairing order and pairing gap decreases slightly from Hg-1212 to CCO, similar to the experimental trend between 3-6 layers, although the small magnitude of the change is challenging within the uncertainty of our numerical approach. Our result for CCO is also in good agreement with the $T_{\mathrm{c}}$ for the mixed Sr/CCO compound (assuming that reflects the $T_{\mathrm{c}}$ of CCO).

The pairing and magnetic orders are shown in Fig. 3D. The qualitative behavior in the mercury-barium cuprates is similar to that in CCO, although here, the hole density is less localized on the Cu atoms, and some fraction goes to buffer atoms (e.g. apical oxygen orbitals). There are other important microscopic differences between the mercury-barium compounds and CCO. For example, in CCO, the magnetic order at half-filling decreases between ambient and 32 GPa pressure, and this is reflected in the more rapid decrease of the local moment close to optimal doping. However, even though the local moment in Hg-1212 decays more slowly than in Hg-1201, the optimal pairing order is larger.

The argon-reduced Hg-1212 structure is similar to the oxy-genated structure but has a larger apical Cu-O distance (by about 0.04 Å). Although the experimental sample corresponds to very low oxygen doping where it is not expected to superconduct, it is still observed to have a $T_c \sim 92$ K [38], leading to speculation about complex charge-transfer behavior in mercury barium cuprates. We find that the undoped magnetic behavior (e.g. exchange couplings and charge distribution) is almost identical in the Hg-1212 (ox) and Hg-1212 (red) structures (see Table S4). Nevertheless, as we dope the reduced structure, we find that holes distribute differently in the reduced and oxidized form, especially near optimal doping ($x_{tot} \sim 0.5$) where the effective Cu doping is smaller in the reduced structure than the oxidized structure, along with a reduction in pairing order (as in the experiment, but much larger in magnitude). The difference in the Hg-1212 (ox) and Hg-1212 (red) electron densities is shown in Fig. 3E: the main difference corresponds to a transfer of charge from the in-plane O $2p$ orbitals in the (red) structure, to the apical O and Hg orbitals in the (ox) structure, leaving the Hg-1212 (ox) with a larger in-plane doping. The sensitivity of pairing order to the charge distribution highlights the need to further investigate the treatment of doping and the charge density. Overall, the complicated behaviour confirms the importance of atomic scale crystal structure in the development of the pairing order in these multicomponent, multilayer cuprates.

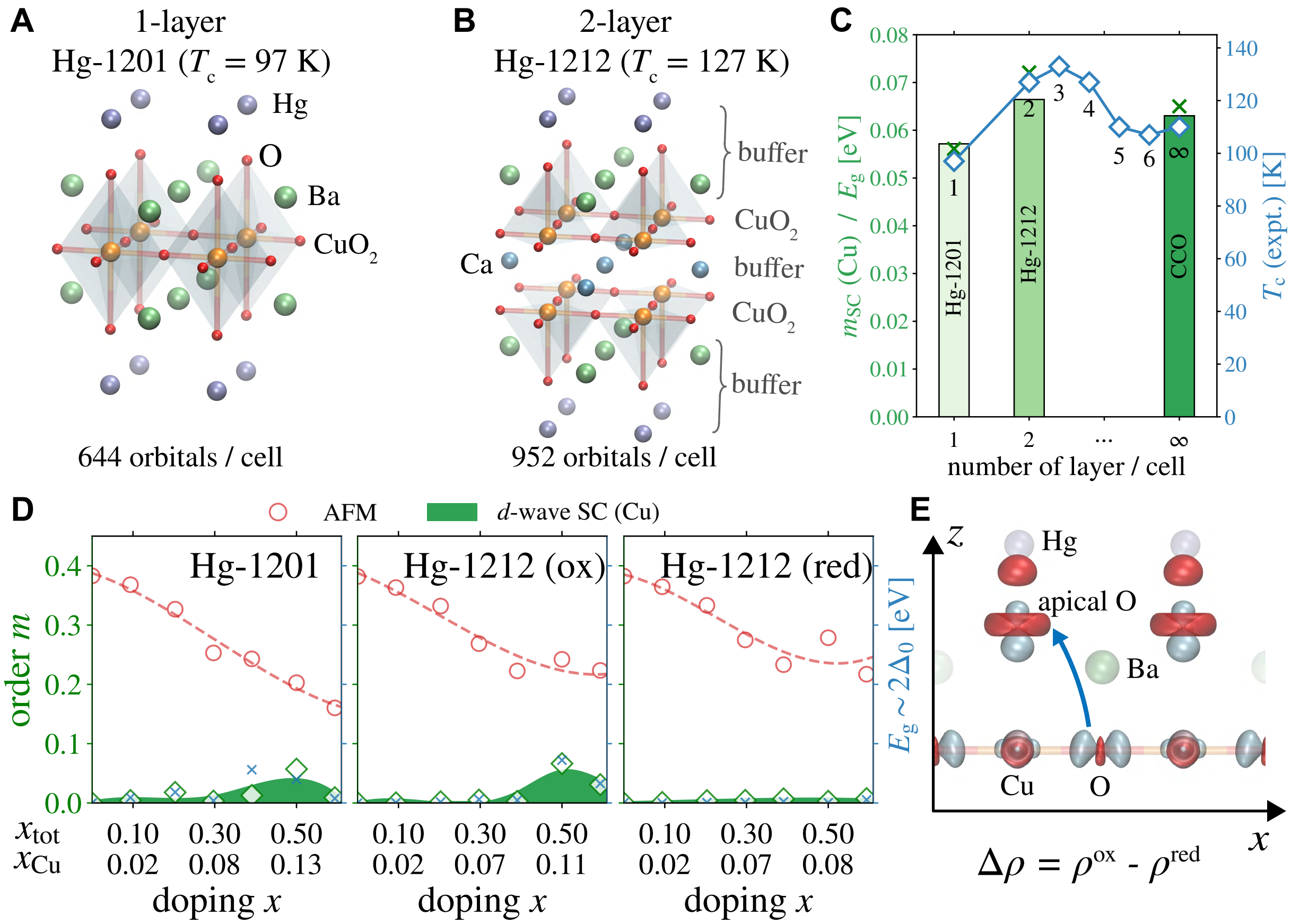

FIG. 3. **Layer effect.** Structures of (A) single-layer Hg-1201 ($HgBa_2CuO_4$) and (B) double-layer Hg-1212 ($HgBa_2CaCu_2O_6$); (C) Comparison between calculated $d$-wave SC orders $m_{SC}$ (bar) and pairing gaps $E_g$ (cross marker) (Hg-1201, Hg-1212, and CCO) and experimental $T_c$ (data from [7, 35]) as a function of the number of layers per cell. (D) AFM, SC orders $m$, and estimated SC gap $E_g$ of different compound structures (Hg-1201, oxidized Hg-1212, and reduced Hg-1212). (E) Electron density difference between oxidized and reduced Hg-1212 at optimal doping (red: increased electron density of oxidized vs. reduced, blue: decreased electron density). The arrow indicates the transfer of charge between reduced and oxidized structures.

## VI. DESCRIPTORS FOR SUPERCONDUCTIVITY IN THE CUPRATES

Our ab initio calculations above capture the correct pressure and layer effects on pairing order across several cuprate structures and compositions. We can therefore interrogate these in silico solutions to identify the features of the electronic structure that most correlate with these trends.

In Fig. 4 we plot maximum pairing order against a variety of descriptors: (i) the magnetic (nearest neighbour Heisenberg) exchange parameter $J$, derived from the same ab initio methodology applied at half-filling following [25], (ii) oxygen hole content at optimal doping ($\Delta n_O$), (iii) the bond order between Cu $3d$-O $2p$ ($\sim \left\langle a^\dagger_{3d_{x^2-y^2}} a_{2p_{x(y)}} \right\rangle^2$, see Eq. (S68)). (i) and

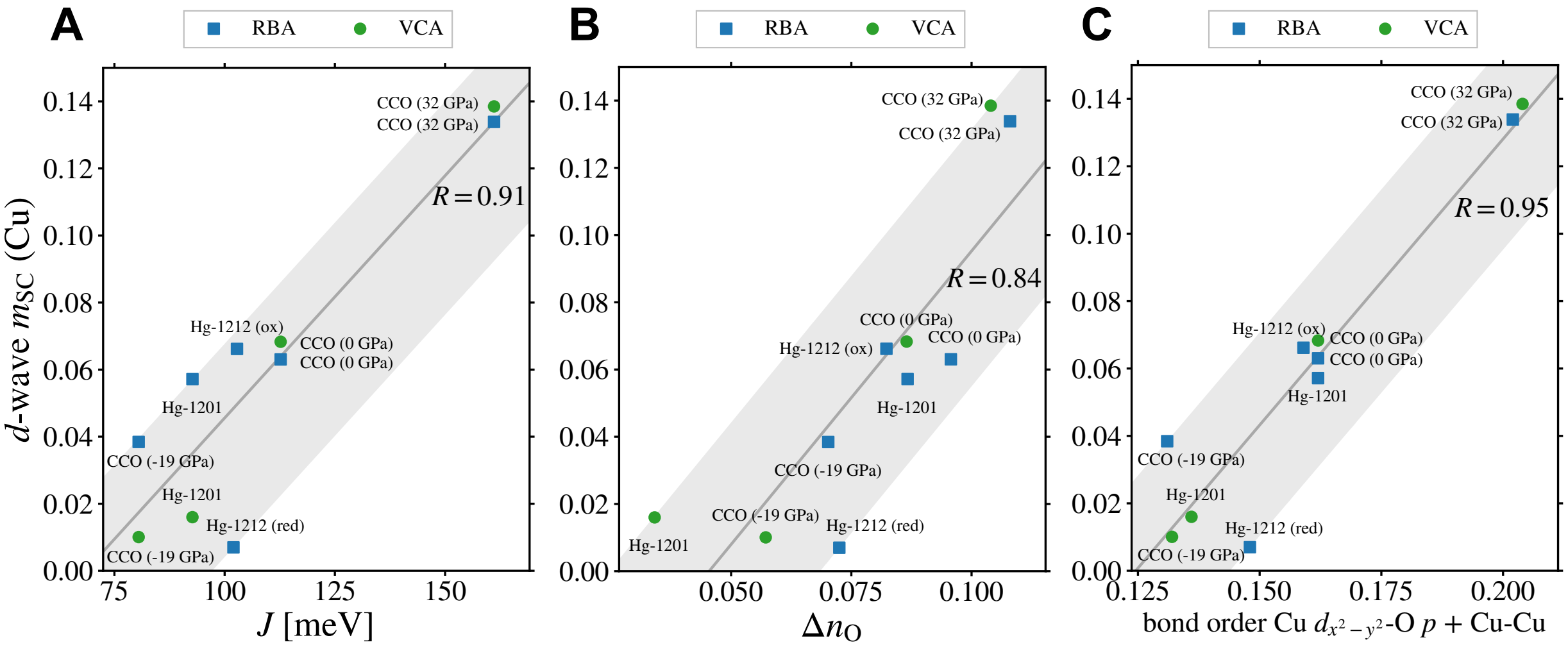

FIG. 4. **Descriptors of superconductivity.** Correlation between superconducting order and selected descriptors: (A) Super-exchange coupling parameter $J$ at half-filling. (B) Average number of holes on oxygen $\Delta n_O$ at optimal doping. (C) Bond order (off-diagonal measure of covalency) at optimal doping. Linear regression $R$ values are also shown. RBA, VCA: rigid band approximation, virtual crystal approximation.

(ii) have previously been invoked as descriptors in the literature (see e.g. [43, 44]); exchange has been associated with cuprate superconductivity since the earliest discussions [45, 46] and its correlation with $T_c$ has attracted much experimental interest [44, 47–49]. Both (ii) and (iii) are related to the charge-transfer gap and covalency of the Cu-O bond, which are commonly discussed in theoretical treatments [21, 50–52] as well as in various experiments [48, 53–56]. However, it should be noted that (ii) and (iii) are related but different probes of these quantities: (ii) measures the diagonal part of the density matrix, while (iii) measures the off-diagonal part. (We note that the $J$ values obtained here are somewhat smaller than have been reported in the experimental literature for some of these compounds [44] (and about 20% smaller than we reported in Ref. [25] due to the different initial density) but we use $J$ values from the same computational approximations as for the rest of this work for internal consistency).

We see that the best qualitative descriptor for the trends in pairing order is the exchange parameter $J$, which captures the general features of both the pressure and layer effect. The correlation between pressure and $J$ is straightforward, as increasing pressure increases the kinetic contribution in the super-exchange mechanism. We discussed the microscopic origins of the layer effect on $J$ in [25]. These are subtle, involving both mean-field (i.e. band structure) [57] and correlated electronic effects [25] with the apical orbitals. However, $J$ does not give the right ordering for Hg-1212 (red) and Hg-1212 (ox), which have almost the same $J$ but very different pairing orders. This is unsurprising, because $J$ is derived from magnetism in the undoped compound, and neglects the material-specific aspects of doping, which we saw were different in Hg-1212 (red) and Hg-1212 (ox). We emphasize that the correlation between $J$ and pairing order appears in our study in a material-specific manner, where both quantities are obtained within the same computational approximation.

Unlike $J$, the oxygen hole content and bond-order at optimal doping are descriptors in the doped state. They both correlate with the pairing order (capturing the pressure effect) and for example, successfully distinguish between Hg-1212 (red) and Hg-1212 (ox), with their different doping dependent electronic structure. The bond-order descriptor has a particularly good average correlation. However, both descriptors do not capture the layer effect. The oxygen hole-content and bond-order reflect mainly the single-particle part of the super-exchange mechanism that gives rise to $J$, and thus do not capture the subtle dependence on the layers.

The strong correlation of the trends in the pairing order with the local descriptors suggests two things. The first is that systematics in pairing can be understood in terms of quantum correlations at a relatively local level. The second is that the physics of superexchange, appropriately modified to account for material-specific doping, is likely behind the trends in pairing. We now examine this possibility.

## VII. MICROSCOPIC ANALYSIS OF FLUCTUATIONS AND PAIRING

To obtain a clearer understanding of the microscopic processes driving pairing, we examine the fluctuations in the correlated quantum impurity wavefunctions $\Psi^{\text{emb}}$ that lead to the SC order. In the one-band Hubbard model, such fluctuation analyses have previously been employed for the 2-particle Green's function [58], and here we devise some time-independent analogs.

Within the CCSD solver, $\Psi^{\text{emb}} = e^{T_1+T_2}|\text{m.f.}\rangle$, where $|\text{m.f.}\rangle$ is a Slater determinant (in the non-SC phase) and a BCS state (in the doped SC phase), where the mean-field is determined by the DMET self-consistency. $T_1$ is an operator containing quadratic fermion terms $(a^\dagger a, a^\dagger a^\dagger)$, while $T_2$ contains quartic fermion terms $(a^\dagger a^\dagger aa, a^\dagger a^\dagger a^\dagger a, a^\dagger a^\dagger a^\dagger a^\dagger, \cdots)$; the exponentiation allows $T_1$ to capture disconnected fluctuations of individual particles, while $T_2$ describes the connected fluctuations of pairs of particles. We can further classify $T_1$ and $T_2$ into components that produce different changes in the particle number: $N$ (normal fluctuations which do not change particle number), and $N \pm 2, \pm 4$ (anomalous fluctuations). Fig. 5A, shows the magnitude of the $T_1$ and $T_2$ components after DMET self-consistency. The fluctuations are largest at intermediate dopings, and the normal fluctuations are larger than the anomalous fluctuations.

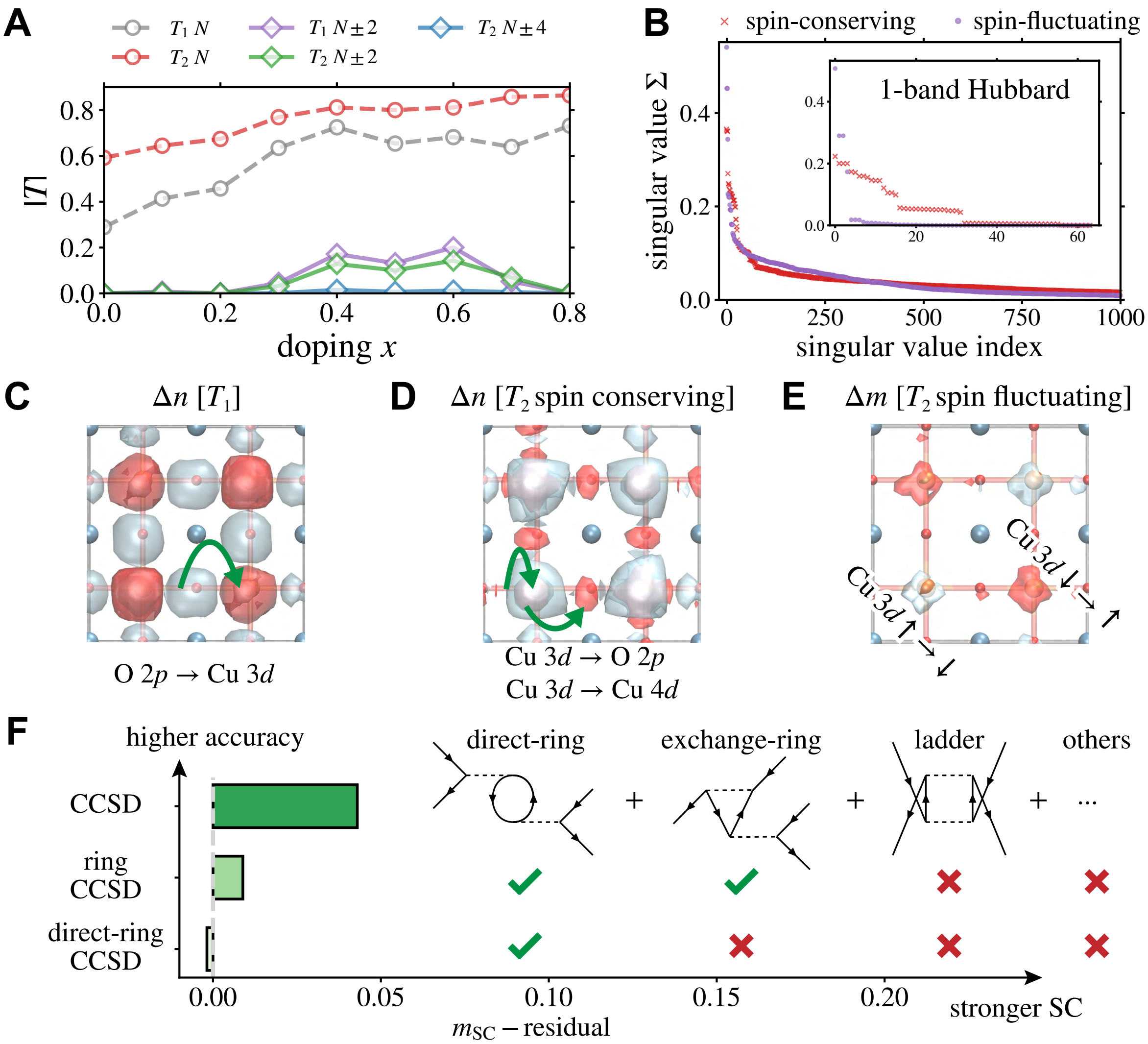

FIG. 5. **Microscopic analysis.** (A) Norm of different blocks of coupled-cluster amplitudes ($N$: normal fluctuations, $N \pm 2$, $N \pm 4$ anomalous fluctuations). (B) Singular values of different $T_2$ decompositions [spin conserving ($\Delta S_z = 0$) or spin fluctuating ($\Delta S_z = \pm 1$)] in an ab initio cuprate calculation (CCO, ambient pressure); inset: $T_2$ decomposition in the one-band Hubbard model plaquette DMET at $U = 6$). (C) Effect of the largest singular mode of $T_1$ on the charge density $n$ (red: $n$ increase; blue: $n$ decrease): charge shifts from oxygen to copper. (D) Effect of the largest singular mode of $T_2$ (spin-conserving decomposition) on the charge density: charge shifts from Cu 3$d$ to empty Cu orbitals (breathing) and to O. (E) Effect of the largest left, right singular modes of $T_2$ (spin-fluctuating decomposition) on the spin density: the moments along the diagonal increase (red)/decrease (blue). (F) SC order (a residual SC associated with the lattice smearing temperature has been subtracted, see Sec. 3.3 of [34]) from different variants of solvers: CCSD includes all 4 classes of diagrams (not limited to those shown), ring CC contains direct and exchange ring terms, whereas direct-ring CC only contains the direct-ring terms.

To show the physical meaning of the $T_1$ fluctuation, we visualize its effect in CCO (ambient pressure) on the charge density in Fig. 5C. The primary components of $T_1$ are excitations between O 2$p$ and Cu 3$d$ orbitals, resulting in a shift of charge density from oxygen to copper. The effect of this correlation-stabilized fluctuation (i.e. it appears only in the presence of non-zero $T_2$) is to make the copper-oxygen bond more covalent than expected in a simple mean-field treatment.

To see the physical meaning of the four-fermion $T_2$ amplitude, we decompose it through a principal component analysis, similar to decompositions of the four-fermion (two-particle) Green's functions [58]. We write $T_2 = \sum_m w_m O_m^\dagger O_m$, and then large weights $w_m$ denote dominant mode. Indeed, within a random phase approximation, the $O_m$ are bosonic operators, and if there is a dominant bosonic mode that is driving the superconducting instability, we would expect this to appear as a large singular value $w_m$. Because $T_2$ preserves $S_z$ symmetry, we can carry out the decomposition into two channels: $\{O_m\}$ such that $[O_m, S_z] = 0$ (spin-conserving channel), or $[O_m, S_z] = \pm 1$ (spin-fluctuation channel). Note that our $T_2$ operator is in the Nambu representation, and the spin-conserving and spin-fluctuating channels contain both particle number preserving and particle number non-conserving $O_m$.

Fig. 5B shows the ordered singular values in the spin-conserving and spin-fluctuating channels for CCO at ambient pressure. (For comparison, we show the same analysis for the DMET plaquette treatment of the 2D pure one-band Hubbard model at $U = 6$; as we are not allowing competition with stripe orders, the next nearest neighbor $t'$ is not critical here to stabilize a superconducting ground state [18]). The largest singular values are in the spin-fluctuating channel. There are similarities between the modes in CCO and in the one-band Hubbard model, in particular, we see 4 large singular values in the $S_z \pm 1$ channel in both cases. However, the distribution is less peaked in the ab initio case, and the difference between the spin-conserving and spin-fluctuating channels is less pronounced. We visualize the dominant modes in Figs. 5D, E. In the spin-conserving channel, the first 4 fluctuations mainly involve charge redistribution from Cu $3d$ to other orbitals, principally, the oxygen $2p$ orbitals and Cu $4d$ (details in Table S5). These types of multi-orbital charge fluctuations, such as "breathing mode" $3d \rightarrow 4d$ excitations [59] and buffer layer excitations beyond the three-band in-plane orbitals of the cuprates, have previously been shown to be essential to capturing material-specific trends in the magnetism in the layer effect [25]. In the spin-fluctuating channel, the first 4 fluctuations flip the spin densities in the Cu $3d_{x^2-y^2}$ orbitals. They are primarily magnon excitation operators, formed from linear combinations of the $3d$ shell spin-flip operators on the Cu atoms with different phases (and are related to the low-lying triplet excitations in model studies of plaquettes [60]).

To connect these different kinds of fluctuations to the generation of pairing, we recompute the pairing order under different approximations in the CC many-body solver. In Fig. 5F we show 4 families of diagrams included in the CCSD solver (shown in the Nambu representation). The direct-ring diagrams lead only to charge fluctuations. The corresponding variant of the CC solver, which only includes the direct-ring diagrams, leads to no pairing order in CCO (ambient pressure) at least within our numerical resolution. (We have subtracted a small residual SC associated with our implementation of finite temperature smearing in the DMET lattice calculation, c.f. Secs. 3.3 and 3.7 of [34]). The remaining diagrams in CCSD introduce spin-fluctuations. For example, including the antisymmetrized, or exchange-ring diagrams (such diagrams vanish in the Hubbard model) adds a sub-class of spin-fluctuations: we see pairing begin to develop. Fig. S8 shows the contribution of only the ladder diagrams (which include different spin-fluctuations) to the SC order, which is of a similar magnitude to that from the full ring CCSD. The bulk of the pairing order emerges only when the coupling between ring and ladder diagrams, as contained in the full set of CCSD diagrams, is included. Overall the direct analysis confirms that short-range spin fluctuations drive the strength of the pairing order. However, in our fully ab initio microscopic picture, additional large fluctuations (such as breathing modes and charge redistribution) associated with covalency are also required, as they generate the energy scale of superexchange. This is an essential difference with low-energy 1-band representations.

## VIII. CONCLUDING REMARKS

In this work, we have demonstrated a fully ab initio many-body simulation strategy that, starting from the material structure, directly approximates the solution of the electronic Schrödinger equation to obtain the superconducting pairing order across a range of geometries and compositions of cuprate materials. We note what is not yet contained within our current treatment: there are no phonons, or long-range spin or charge fluctuations (at distances much larger than the computational cell) and long wavelength orders, the treatment of doping does not include explicit dopants and the associated structural relaxation and disorder, nor are the solvers and representations numerically exact. Perhaps the simplest, but amongst the numerically most significant approximations, is the lack of charge self-consistency, which makes our calculations depend on the chosen starting point. (For a more detailed analysis of outstanding limitations, see Sec. 4 of [34]). Consequently, there remain important differences between the results of the simulations and observations in real materials. Nonetheless, our goal is not to reproduce all aspects of cuprate ground-state physics, but rather to capture some of the observed material trends. In this regard, we find that we obtain two trends in these materials: an increase in pairing order and pairing gap as a function of intralayer pressure; and an increase (and then decrease) in maximum pairing order and pairing gap as a function of the number of stacked copper-oxygen layers. These trends are highly reminiscent of similar trends that are experimentally seen in the superconducting critical temperatures. That these trends correctly appear indicates that the physics and numerical aspects of the calculation likely contain important and relevant ingredients to describe superconducting pairing in a material-specific manner across a range of cuprate compounds.

Detailed analysis of our calculations supports some long-standing proposals for the driving force for pairing in cuprates, but also provides new insights. Superexchange, suitably defined, correlates well with maximum pairing order, and short-range spin fluctuations, mainly on the copper atoms, drive the pairing. However, the ab initio picture of the fluctuations is richer than that in simplified models, because multi-orbital effects associated with covalency are needed to facilitate the spin fluctuations. Such multi-orbital processes are key to material-specific trends.

There is much room to systematically improve the computations in this work in the future. At the same time, we see our material-specific modeling of the superconducting ground-state as the starting point for a material-driven understanding of the full phase diagram; our atomically and orbitally resolved diagrammatic and fluctuation analysis as a route to elucidating the microscopic mechanisms underlying the phases; and the identification of central descriptors as aiding the computational search for new high-temperature superconducting materials. Importantly, our work shows that targetting a material-specific understanding of superconductivity in the cuprates is now a realistic goal through direct ab initio computation.

**Code Availability.** Codes used in this work can be found in the following repositories. The libDMET repository is at https://github.com/gkclab/libdmet_preview [61]. The Block2 code repository is at https://github.com/block-hczhai/block2-preview. PySCF is available from https://pyscf.org.

**Acknowledgements.** We thank David Reichman, Andrew Millis, Tianyu Zhu, Linqing Peng, Yuan Li, Patrick Lee, Steve White, Dingshun Lv, Lei Wang, Hong Jiang, Nai-Chang Yeh and Sandeep Sharma for helpful discussions.

This work was primarily supported by the US Department of Energy, Office of Science, via grant no. DE-SC0018140. G.K.-L.C. is a Simons Investigator in Physics. Calculations were performed using the facilities of National Energy Research Scientific Computing Center (NERSC), a U.S. Department of Energy Office of Science User Facility located at Lawrence Berkeley National Laboratory, under NERSC award ERCAP0023924, and in the Resnick High Performance Computing Center, supported by the Resnick Sustainability Institute at Caltech. We thank M. Graf and C. Mewes for arranging for computational time on NERSC. This work was partially supported by the National Science Foundation under Award No. CHE-1848369 (H.-Z.Y., development of Gaussian basis) and the Air Force Office of Scientific Research under Award No. FA9550-18-1-0095 (R. K., correlation potential fitting). L. L. is a Simons Investigator. J.T. acknowledges funding by the Deutsche Forschungsgemeinschaft (DFG, German Research Foundation) through DFG-495279997.

**Supplementary Information.** Theory and implementation, Figures S1-S12, Tables S1-S5, References.

---

# Supplementary Materials for "Ab initio quantum many-body description of superconducting trends in the cuprates"

Zhi-Hao Cui [1†§], Junjie Yang [1], Johannes Tölle [1], Hong-Zhou Ye [2],
Shunyue Yuan [3], Huanchen Zhai [1], Gunhee Park [3], Raehyun Kim [4], Xing Zhang [1],
Lin Lin [4,5], Timothy C. Berkelbach [2], Garnet Kin-Lic Chan [1*]

[1]Division of Chemistry and Chemical Engineering, California Institute of Technology,
Pasadena, California 91125, USA
[2]Department of Chemistry, Columbia University, New York, New York 10027, USA
[3]Division of Engineering and Applied Science, California Institute of Technology,
Pasadena, California 91125, USA
[4]Department of Mathematics, University of California, Berkeley, California 94720, USA
[5]Computational Research Division, Lawrence Berkeley National Laboratory,
Berkeley, California 94720, USA

[†]E-mail: zhcui0408@gmail.com *E-mail: gkc1000@gmail.com

[§]Present address: Department of Chemistry, Columbia University, New York, New York 10027, USA.

# 1 Methods

In the sections below, we describe specifics of the DMET formalism, quantum chemistry solvers, and analysis as used in the main text. In the first two cases, we focus only on those aspects new to ab initio DMET applications to superconducting ground-states; we do not recount the entire ab initio DMET quantum chemistry technology. The ab initio DMET implementation is described in detail in Ref. [62] and the SI of Ref. [25].

## 1.1 Quantum chemistry formalism for superconductivity

### 1.1.1 Recap of BCS theory, the Nambu representation, and pairing symmetry

Bardeen–Cooper–Schrieffer (BCS) theory provides a mean-field treatment of superconductivity. In this theory, superconducting order corresponds to a non-zero expectation value of $\langle a^\dagger a^\dagger \rangle$ and $\langle aa \rangle$. In other words, particle number is not conserved in the BCS ground-state, and superconductivity corresponds to a broken particle-number symmetry ($U(1)$ symmetry).

The original BCS formulation uses a one-band picture and a simplified two-electron Hamiltonian,

$$\hat{H}^{\text{simplified}} = \sum_{\mathbf{k}\sigma} (\varepsilon_{\mathbf{k}} - \mu) a^\dagger_{\mathbf{k}\sigma} a_{\mathbf{k}\sigma} - \sum_{\mathbf{k}\mathbf{k}'} V_{\mathbf{k}\mathbf{k}'} a^\dagger_{\mathbf{k}\alpha} a^\dagger_{-\mathbf{k}\beta} a_{-\mathbf{k}'\beta} a_{\mathbf{k}'\alpha}, \tag{S1}$$

where the one-electron term involves orbital energies $\varepsilon_{\mathbf{k}}$ and a chemical potential $\mu$. Note the negative sign before the two-electron terms $V_{\mathbf{k}\mathbf{k}'}$, i.e. the effective electron-electron interaction is assumed to be attractive. Using Wick's theorem, the two-electron part of the Hamiltonian is approximated in the BCS treatment by a mean-field,

$$-V_{\mathbf{k}\mathbf{k}'} a^\dagger_{\mathbf{k}\alpha} a^\dagger_{-\mathbf{k}\beta} a_{-\mathbf{k}'\beta} a_{\mathbf{k}'\alpha} \approx -V_{\mathbf{k}\mathbf{k}'} \left( \left\langle a^\dagger_{-\mathbf{k}\beta} a_{-\mathbf{k}'\beta} \right\rangle a^\dagger_{\mathbf{k}\alpha} a_{\mathbf{k}'\alpha} - \left\langle a^\dagger_{\mathbf{k}\alpha} a_{-\mathbf{k}'\beta} \right\rangle a^\dagger_{-\mathbf{k}\beta} a_{\mathbf{k}'\alpha} + a^\dagger_{\mathbf{k}\alpha} a^\dagger_{-\mathbf{k}\beta} \left\langle a_{-\mathbf{k}'\beta} a_{\mathbf{k}'\alpha} \right\rangle \right), \tag{S2}$$

where the first two terms are the normal Coulomb ($J_{\mathbf{k}}$) and exchange ($K_{\mathbf{k}}$) potentials, while the last term is the pairing potential. Writing the superconducting pairing potential as

$$\Delta_{\mathbf{k}} = \sum_{\mathbf{k}'} -V_{\mathbf{k}\mathbf{k}'} \left\langle a_{-\mathbf{k}'\beta} a_{\mathbf{k}'\alpha} \right\rangle. \tag{S3}$$

and absorbing the $J_{\mathbf{k}}$ and $K_{\mathbf{k}}$ terms (by defining $\tilde{\varepsilon}_{\mathbf{k}} \equiv \varepsilon_{\mathbf{k}} + J_{\mathbf{k}} - K_{\mathbf{k}}$), we obtain the mean-field BCS Hamiltonian for superconductivity,

$$\hat{H}^{\text{BCS}} = \sum_{\mathbf{k}} \left[ \sum_{\sigma} (\tilde{\varepsilon}_{\mathbf{k}} - \mu) a^\dagger_{\mathbf{k}\sigma} a_{\mathbf{k}\sigma} + \left( \Delta_{\mathbf{k}} a^\dagger_{\mathbf{k}\alpha} a^\dagger_{-\mathbf{k}\beta} + \text{H.c.} \right) \right] + \text{const.}. \tag{S4}$$

To obtain the ground-state, we diagonalize this quadratic Hamiltonian using a Bogoliubov transformation [63, 64],

$$\begin{bmatrix} a_{\mathbf{k}\alpha} \\ a^\dagger_{-\mathbf{k}\beta} \end{bmatrix} = \begin{bmatrix} u_{\mathbf{k}} & v_{\mathbf{k}} \\ -v_{\mathbf{k}} & u_{\mathbf{k}} \end{bmatrix} \begin{bmatrix} p_{\mathbf{k}} \\ q^\dagger_{\mathbf{k}} \end{bmatrix}. \tag{S5}$$

The new quasiparticles ($p$ and $q$) mix particles and holes, and $u$ and $v$ are coefficients determined by the diagonalization and normalization conditions. In the new quasiparticle basis, the Hamiltonian becomes,

$$\hat{H}^{\text{BCS}} = \sum_{\mathbf{k}} E_{\mathbf{k}} (p^\dagger_{\mathbf{k}} p_{\mathbf{k}} + q^\dagger_{\mathbf{k}} q_{\mathbf{k}}) + \sum_{\mathbf{k}} (\tilde{\varepsilon}_{\mathbf{k}} - \mu - E_{\mathbf{k}}) + \text{const.}, \tag{S6}$$

where

$$E_{\mathbf{k}} = \sqrt{(\tilde{\varepsilon}_{\mathbf{k}} - \mu)^2 + \Delta^2_{\mathbf{k}}}. \tag{S7}$$

The order parameter $\Delta_{\mathbf{k}}$ is the also the gap function of the elementary excitations since $E_{\mathbf{k}} = |\Delta_{\mathbf{k}}|$ at the Fermi level ($\tilde{\varepsilon} = \mu$). The non-zero pairing also lowers the ground-state energy relative to the original fermion sea for any non-zero pairing. Thus an arbitrarily small attraction leads to the BCS ground-state.

In the current work, the BCS formalism enters the DMET calculation in an auxiliary way. In the DMET procedure, we self-consistently solve an interacting quantum impurity problem (with an interacting Hamiltonian $\hat{H}^{\mathrm{emb}}$, and an auxiliary mean-field problem, with Hamiltonian $\hat{H}^{\mathrm{latt}}$. $\hat{H}^{\mathrm{latt}} = \hat{\varepsilon} + \hat{J} + \hat{K} + \hat{\Delta}$ can be viewed as a type of BCS Hamiltonian where the one-body operator $\hat{\varepsilon}$, $\hat{J}$ and $\hat{K}$ come from a first principles calculation (with many bands) and $\hat{\Delta}$ is determined not by Wick's theorem, but by the DMET self-consistency with the interacting quantum impurity problem. In other words, finding the ground-state of $H^{\mathrm{latt}}$ corresponds to finding a BCS ground-state in a many-band picture, using a pair-potential supplied by the DMET self-consistency. (Note, however, that observables in DMET are computed using the ground-state of $H^{\mathrm{emb}}$, not $H^{\mathrm{latt}}$).

As mentioned in the main text, breaking particle number symmetry is inconvenient in quantum chemistry formulations. A formally equivalent alternative to working with the BCS ground-state and Hamiltonian is to redefine the vacuum so that the particle-number breaking terms become normal ones. We write

$$
\begin{aligned}
\hat{H}^{\mathrm{BCS}} &= \sum_{\mathbf{k}} \begin{bmatrix} a^{\dagger}_{\mathbf{k}\alpha} & a_{-\mathbf{k}\beta} \end{bmatrix} \begin{bmatrix} \tilde{\varepsilon}_{\mathbf{k}} - \mu & \Delta_{\mathbf{k}} \\ \Delta^{\dagger}_{\mathbf{k}} & -\tilde{\varepsilon}_{\mathbf{k}} + \mu \end{bmatrix} \begin{bmatrix} a_{\mathbf{k}\alpha} \\ a^{\dagger}_{-\mathbf{k}\beta} \end{bmatrix} \\
&= \sum_{\mathbf{k}} \begin{bmatrix} c^{\dagger}_{\mathbf{k}\alpha} & c^{\dagger}_{\mathbf{k}\beta} \end{bmatrix} \begin{bmatrix} \tilde{\varepsilon}_{\mathbf{k}} - \mu & \Delta_{\mathbf{k}} \\ \Delta^{\dagger}_{\mathbf{k}} & -\tilde{\varepsilon}_{\mathbf{k}} + \mu \end{bmatrix} \begin{bmatrix} c_{\mathbf{k}\alpha} \\ c_{\mathbf{k}\beta} \end{bmatrix}.
\end{aligned} \tag{S8}
$$

The second line is called the *Nambu representation*[33, 65], which effectively defines the following partial particle-hole (p-h) transformation,

$$
\begin{aligned}
c^{\dagger}_{\mathbf{k}\alpha} &= a^{\dagger}_{\mathbf{k}\alpha} \\
c_{\mathbf{k}\beta} &= a^{\dagger}_{-\mathbf{k}\beta},
\end{aligned} \tag{S9}
$$

The new vacuum of $\beta$ particles is a ferromagnetic state where all the $\mathbf{k}$ states are occupied by spin-down electrons, such that any further creation of $\beta$ electrons will destroy the state. In the Nambu representation, the particle-number symmetry ($c^{\dagger}c$) becomes conserved at the expense of spin symmetry $S_z$ breaking. Diagonalizing the quadratic Hamiltonian gives exactly the Bogoliubov coefficients in Eq. (S5). In quantum chemistry language, in the Nambu representation the Hamiltonian has a spin-coupling block $\Delta$ which mixes the two flavors of spin, similar to a spin-orbit coupling term. Therefore, we call this formulation a *generalized spin orbital* (GSO) formalism, which connects the superconducting problem to the treatment of spin-orbit coupling in quantum chemistry. We use the Nambu representation when formulating the *ab initio* quantum embedding procedure using GSOs in the next section.

We note that in the pair potential in Eq. (S3) we have assumed *singlet pairing* (strictly $S_z = 0$ pairing), i.e. the electron pair is composed of different spins with order parameter $\langle a_{\alpha} a_{\beta} \rangle$. It is also possible to have triplet pairing, where $\langle a_{\alpha} a_{\alpha} \rangle$ is non-zero. In that case, the Nambu representation can still be used, but the resulting Hamiltonian does not have the simple interpretation of generalized spin mixing (e.g., see [66, 67]). The number of orbitals in the Hamiltonian becomes $4N_{\mathrm{orb}}$ [$N_{\mathrm{orb}}$ is the number of sites in the lattice] (instead of $2N_{\mathrm{orb}}$ in the pure singlet pairing case). Triplet pairing is associated with $p$ or $f$ symmetry in the superconducting order parameter. Since the cuprate pairing is known to be $d$-wave, we do not consider triplet pairing further in this work.

We next discuss pairing symmetry [68]. We can view the order parameter $\Delta$ as the amplitude of a two-electron wavefunction. In the absence of spin-orbit coupling, a two-electron wavefunction can be decomposed into spatial and spin components,

$$
\Psi(\mathbf{r}_1, \sigma_1, \mathbf{r}_2, \sigma_2) = \Delta(\mathbf{R}, \mathbf{r}) \chi(\sigma_1, \sigma_2), \tag{S10}
$$

where $\mathbf{R} = \mathbf{r}_1 + \mathbf{r}_2$ is the center of mass and $\mathbf{r} = \mathbf{r}_1 - \mathbf{r}_2$ is the relative coordinate. Assuming singlet pairing ($\langle \hat{S}^2 \rangle = 0$), the spin component $\chi(\sigma_1, \sigma_2) = -\chi(\sigma_2, \sigma_1)$ is anti-symmetric, and since the overall wavefunction is anti-symmetric, the spatial part $\Delta$ must be symmetric. Assuming the Hamiltonian is spatially rotationally invariant, we can classify $\Delta$ by the angular momentum quantum number $l$. The symmetric $\Delta$ then has even integer $l$, i.e. $s$ ($l = 0$), $d$ ($l = 2$), etc. The corresponding superconductor order parameters are called $s$-wave, $d$-wave, etc. respectively. In real materials, there is no continuous rotational invariance due to the lattice. One can, however, still classify the pairing symmetry using $l$ by considering how the irreps reduce from the continuous rotational group to the point group associated with the cell. We note that in our calculations, the simultaneous presence of AFM order and SC order means that the order parameter breaks $\hat{S}^2$ symmetry, and this allows $p$-wave components to appear. We discuss the precise definition of the order parameters further in Sec. 1.5.

#### 1.1.2 Ab initio superconducting Hamiltonians

In this section, we present useful formulae to build the Hamiltonians and matrix elements used in the DMET calculations, accounting for the additional steps of the Nambu transformation and subsequent normal ordering. As discussed in detail in Ref. [62], formally, we start with a second-quantized Hamiltonian in a crystal atomic orbital basis in the bulk lattice; the crystal atomic orbital basis gives rise to a set of one- and two-electron integrals labelled by the basis indices e.g. $p, q, r, s$, and $\mathbf{k}$-point indices $\mathbf{k}_p, \mathbf{k}_q, \ldots$. We will use $H_1$ and $V_2$ to generically denote one-electron and two-electron integral tensors (for any Hamiltonian). DMET works with an embedding Hamiltonian $\hat{H}^{\mathrm{emb}}$ and a mean-field bulk Hamiltonian $\hat{H}^{\mathrm{latt}}$ and the matrix elements of these Hamiltonians are derived from the bulk $H_1$ and $V_2$ integrals through integral transformations and contractions with density matrices [62]. In the current work, in addition we need to perform the partial p-h transformation of the Nambu formalism,

$$\begin{aligned} c^\dagger_{i\alpha} &= a^\dagger_{i\alpha} \\ c_{j\beta} &= a^\dagger_{j\beta}. \end{aligned} \tag{S11}$$

followed by normal ordering to obtain integrals in the quasiparticle representation, where we use calligraphic letters ($\mathcal{H}_1$, $\mathcal{V}_2$, etc) to denote the transformed quantities. After the Nambu transformation, all Hamiltonian components are particle-number conserving as used in generalized spin orbital quantum chemistry treatments.

- Overlap matrix and orbital coefficients

  The partial p-h transform of the basis overlap matrix $S$ (which may be different from the identity in the crystal atomic orbital basis) or orbital coefficients $C$ reads,

$$S^{\mathbf{k}} \to \mathcal{S}^{\mathbf{k}} \equiv \begin{bmatrix} S^{\mathbf{k}} & 0 \\ 0 & S^{\mathbf{k}} \end{bmatrix}, \tag{S12}$$

$$C^{\mathbf{k}} = \left[ C^{\alpha\mathbf{k}}, C^{\beta\mathbf{k}} \right] \to \mathcal{C}^{\mathbf{k}} \equiv \begin{bmatrix} C^{\alpha\mathbf{k}} & 0 \\ 0 & C^{\beta\mathbf{k}} \end{bmatrix}. \tag{S13}$$

- One-body Hamiltonian

  The partial p-h transform of a one-body Hamiltonian, $H_1^{\mathbf{k}}$, is

$$H_1^{\mathbf{k}} = \left[ h^{\alpha\mathbf{k}}, h^{\beta\mathbf{k}}, \Delta^{\alpha\beta\mathbf{k}} \right] \to \mathcal{H}_1^{\mathbf{k}} \equiv \begin{bmatrix} h^{\alpha\mathbf{k}} & \Delta^{\alpha\beta\mathbf{k}} \\ \Delta^{\alpha\beta\mathbf{k}\dagger} & -h^{\beta\mathbf{k}} \end{bmatrix} + \mathcal{H}_0, \tag{S14}$$

  where the energy constant $\mathcal{H}_0 = \sum_{\mathbf{k}} \mathrm{Tr}\left( h^{\beta\mathbf{k}} S^{\mathbf{k},-1} \right)$ comes from the normal ordering (change of vacuum).

- One-body density matrix

  The partial p-h transformation of a one-body reduced density matrix $\gamma_1^{\mathbf{k}}$ reads,

$$\gamma_1^{\mathbf{k}} = \left[ \gamma^{\alpha\mathbf{k}}, \gamma^{\beta\mathbf{k}}, \kappa^{\alpha\beta\mathbf{k}} \right] \to \mathcal{D}^{\mathbf{k}} \equiv \begin{bmatrix} \gamma^{\alpha\mathbf{k}} & \kappa^{\alpha\beta\mathbf{k}} \\ \kappa^{\alpha\beta\mathbf{k}\dagger} & S^{\mathbf{k},-1} - \gamma^{\beta\mathbf{k}} \end{bmatrix}. \tag{S15}$$

  $\mathcal{D}$ will serve as the *generalized density matrix* for constructing the embedding orbitals $\mathcal{C}$ of DMET. The physical number of electrons (not the number of quasiparticles) can be computed as,

$$\begin{aligned} N^{\mathrm{elec}} &= \frac{1}{N_{\mathbf{k}}} \sum_{\mathbf{k}} \mathrm{Tr}\left\{ \gamma^{\alpha\mathbf{k}} S^{\mathbf{k}} \right\} + \mathrm{Tr}\left\{ \gamma^{\beta\mathbf{k}} S^{\mathbf{k}} \right\} \\ &= \frac{1}{N_{\mathbf{k}}} \sum_{\mathbf{k}} \mathrm{Tr}\left( \mathcal{D}^{\mathbf{k}} \mathcal{S}^{\mathbf{k}} \right)^{\alpha\alpha} + \mathrm{Tr}\left( I - \mathcal{D}^{\mathbf{k}} \mathcal{S}^{\mathbf{k}} \right)^{\beta\beta} \end{aligned} \tag{S16}$$

- Two-body Hamiltonian

  The partial p-h transform of an electron repulsion integral (ERI), $V_2$, is

$$V_2 \to \mathcal{V}_2 \equiv \begin{bmatrix} V & -V \\ -V^\dagger & V \end{bmatrix} + \mathcal{V}_1 + \mathcal{V}_0, \tag{S17}$$

  where

$$\mathcal{V}_1 = \begin{bmatrix} v^J & 0 \\ 0 & v^K - v^J \end{bmatrix} \tag{S18}$$

  with

$$\begin{aligned} v^J_{rs} &= \sum_{pq} S^{-1}_{qp} V_{pqrs}, \\ v^K_{ps} &= \sum_{qr} S^{-1}_{qr} V_{pqrs}, \end{aligned} \tag{S19}$$

  and

$$\mathcal{V}_0 = \frac{1}{2} \mathrm{Tr}\big[(v_J - v_K) S^{-1}\big]. \tag{S20}$$

- Two-body Hamiltonian from density fitting integrals

  In realistic materials calculations, it is usually infeasible to store the entire two-electron integral tensor $V_2$ for the whole lattice. Instead, as discussed in Ref. [62], we use density fitting to obtain a low-rank decomposed $V_2$ expressed in terms of 3-center (crystal) atomic orbital (AO) integrals $W^{\mathbf{k}_p\mathbf{k}_q}_{Lpq}$, where $L$ is an index in auxiliary density-fitting basis, and $p(q)$ are indices in the crystal AO computational basis (see Sec. 2.3 for details), and reconstruct the relevant $V_2$ integrals when needed. When constructing the $V_2$ or Fock matrix of the embedded problem, we never explicitly build the lattice $V_2$ (we only need to build $\mathcal{V}_1$ and $\mathcal{V}_0$). The relevant formulae are summarized in the following.

  To construct $\mathcal{V}_1$ using density fitting, we use

$$\mathcal{V}^{\mathbf{k}}_1 = \begin{bmatrix} v^{J\mathbf{k}} & 0 \\ 0 & v^{K\mathbf{k}} - v^{J\mathbf{k}} \end{bmatrix}, \tag{S21}$$

  where $\mathbf{k}$ denotes the $\mathbf{k}$-point of interest, and $v^{J\mathbf{k}}$ is computed as

$$\rho_L = \frac{1}{N_{\mathbf{k}}} \sum_{\mathbf{k}pq} S^{\mathbf{k},-1}_{qp} W^{\mathbf{k}\mathbf{k}}_{Lpq}, \tag{S22}$$

$$v^{J\mathbf{k}}_{rs} = \sum_L \rho_L W^{\mathbf{k}\mathbf{k}}_{Lrs}. \tag{S23}$$

  Similarly, $v^{K\mathbf{k}}$ is evaluated through

$$X^{\mathbf{k}_q\mathbf{k}_p}_{Lqs} = \sum_r S^{\mathbf{k}_q,-1}_{qr} W^{\mathbf{k}_q\mathbf{k}_p}_{Lrs}, \tag{S24}$$

$$v^{K\mathbf{k}_p}_{ps} = \frac{1}{N_{\mathbf{k}}} \sum_{\mathbf{k}_q Lq} W^{\mathbf{k}_q\mathbf{k}_p *}_{Lqp} X^{\mathbf{k}_q\mathbf{k}_p}_{Lqs}. \tag{S25}$$

  The constant term is,

$$\mathcal{V}_0 = \frac{1}{2N_{\mathbf{k}}} \sum_{\mathbf{k}} \mathrm{Tr}\Big[(v^{J\mathbf{k}} - v^{K\mathbf{k}}) S^{\mathbf{k},-1}\Big]. \tag{S26}$$

- Embedding Hamiltonian

  The embedding orbitals (impurity $\mathcal{I}$ + bath $\mathcal{B}$) are constructed via the SVD of the off-diagonal block of the generalized density matrix $\mathcal{D}^{\mathbf{R}\neq\mathbf{0}}$ [62],

$$\mathcal{D}^{\mathbf{R}\neq\mathbf{0}} = \mathcal{B}^{\mathbf{R}\neq\mathbf{0}}\Lambda V^{\dagger}, \tag{S27}$$

  where we have taken the first cell ($\mathbf{R} = \mathbf{0}$) as impurity. The embedding space (impurity + bath) are then spanned by the union of $\mathcal{I}$ and $\mathcal{B}$,

$$\mathcal{C}^{\mathbf{k}} = \sum_{\mathbf{R}} \mathrm{e}^{-i\mathbf{k}\cdot\mathbf{R}}\mathcal{C}^{\mathbf{R}} = \sum_{\mathbf{R}} \mathrm{e}^{-i\mathbf{k}\cdot\mathbf{R}} \begin{bmatrix} \mathcal{I}^{\mathbf{R}=\mathbf{0}} & \mathbf{0} \\ \mathbf{0} & \mathcal{B}^{\mathbf{R}\neq\mathbf{0}} \end{bmatrix}. \tag{S28}$$

  The construction of the embedding one-body Hamiltonian reads as,

$$\mathcal{H}^{\mathrm{emb}}_{ij} = \frac{1}{N_{\mathbf{k}}}\sum_{\mathbf{k}} \mathcal{C}^{\mathbf{k}\dagger}_{ip}\mathcal{F}^{\mathbf{k}}_{pq}\mathcal{C}^{\mathbf{k}}_{qj} - \left(\mathcal{J}_{ij}[\mathcal{D}^{\mathrm{emb}}] - \mathcal{K}_{ij}[\mathcal{D}^{\mathrm{emb}}]\right) \tag{S29}$$

  where the generalized Fock matrix $\mathcal{F}$ is projected using the embedding basis $\mathcal{C}$ and subtracted by the double-counting embedding Coulomb ($\mathcal{J}$) and exchange ($\mathcal{K}$) contributions. These would require both the embedding density matrix,

$$\mathcal{D}^{\mathrm{emb}} = \frac{1}{N_{\mathbf{k}}}\sum_{\mathbf{k}} \mathcal{C}^{\mathbf{k}\dagger}\mathcal{S}^{\mathbf{k}}\mathcal{D}^{\mathbf{k}}\mathcal{S}^{\mathbf{k}}\mathcal{C}^{\mathbf{k}}, \tag{S30}$$

  and the embedding ERI,

$$\begin{aligned} \mathcal{V}^{\mathrm{emb}}_{ijkl} = \frac{1}{N_{\mathbf{k}}}\sum_{\mathbf{k}_L L} \Big( & W^{\alpha\alpha\mathbf{k}_L\mathbf{00}*}_{Lij} W^{\alpha\alpha\mathbf{k}_L\mathbf{00}}_{Lkl} + W^{\beta\beta\mathbf{k}_L\mathbf{00}*}_{Lij} W^{\beta\beta\mathbf{k}_L\mathbf{00}}_{Lkl} \\ & - W^{\alpha\alpha\mathbf{k}_L\mathbf{00}*}_{Lij} W^{\beta\beta\mathbf{k}_L\mathbf{00}}_{Lkl} - W^{\beta\beta\mathbf{k}_L\mathbf{00}*}_{Lij} W^{\alpha\alpha\mathbf{k}_L\mathbf{00}}_{Lkl} \Big), \end{aligned} \tag{S31}$$

  where the reference cell 3-center embedding integral $W$ is calculated as,

$$W^{\sigma\tau\mathbf{k}_L\mathbf{00}}_{Lij} = \frac{1}{N_{\mathbf{k}}}\sum_{\mathbf{k}_p\mathbf{k}_q}{}^{'} \sum_{pq} \mathcal{C}^{\sigma\mathbf{k}_p\dagger}_{ip} W^{\mathbf{k}_p\mathbf{k}_q}_{Lpq}\mathcal{C}^{\tau\mathbf{k}_q}_{qj}, \tag{S32}$$

  where $\mathcal{C}^{\sigma}$ is the coefficient matrix of embedding orbitals with spin $\sigma = \{\alpha, \beta\}$ and the $'$ limits the summation through momentum conservation $\mathbf{k}_L = \mathbf{k}_p - \mathbf{k}_q + n\mathbf{b}$.

- Coulomb and exchange potential

  The Coulomb $\mathcal{J}$ and exchange $\mathcal{K}$ potential in the quasiparticle representation is defined via a contraction between $\mathcal{V}_2$ and $\mathcal{D}_1$,

$$\mathcal{J}_{rs} = \sum_{pq}\mathcal{D}_{qp}\mathcal{V}_{pqrs}, \tag{S33}$$

$$\mathcal{K}_{ps} = \sum_{qr}\mathcal{D}_{qr}\mathcal{V}_{pqrs}. \tag{S34}$$

  Substituting Eq. (S15) and (S17), we have

$$\begin{aligned} \mathcal{J}^{\alpha}_{rs} &= \sum_{pq}\mathcal{D}^{\alpha}_{qp}\mathcal{V}^{\alpha\alpha}_{pqrs} + \mathcal{D}^{\beta}_{qp}\mathcal{V}^{\beta\alpha}_{pqrs} \\ &= \sum_{pq}\mathcal{D}^{\alpha}_{qp}V_{pqrs} - \mathcal{D}^{\beta}_{qp}V_{pqrs} \\ &= \sum_{pq}\gamma^{\alpha}_{pq}V_{pqrs} + \gamma^{\beta}_{pq}V_{pqrs} - S^{-1}_{pq}V_{pqrs} \\ &= \left(\mathcal{T}(J^{\alpha})\right)_{rs} - v^{J}_{rs}, \end{aligned} \tag{S35}$$

$$
\begin{aligned}
\mathcal{J}^{\beta}_{rs} &= \sum_{pq} \mathcal{D}^{\beta}_{qp}\mathcal{V}^{\beta\beta}_{pqrs} + \mathcal{D}^{\alpha}_{qp}\mathcal{V}^{\alpha\beta}_{pqrs} \\
&= \sum_{pq} \mathcal{D}^{\beta}_{qp} V_{pqrs} - \mathcal{D}^{\alpha}_{qp} V_{pqrs} \\
&= \sum_{pq} -(\gamma^{\beta}_{pq} V_{pqrs} + \gamma^{\alpha}_{pq} V_{pqrs}) + S^{-1}_{pq} V_{pqrs} \\
&= \left(\mathcal{T}(J^{\beta})\right)_{rs} + v^{J}_{rs}.
\end{aligned}
\tag{S36}
$$

where we have defined a mapping operator $\mathcal{T}$ for a matrix $A$ to the Nambu representation: $\mathcal{T}(A^{\alpha}) = A^{\alpha}$ and $\mathcal{T}(A^{\beta}) = -A^{\beta}$. Note that this result is different from that obtained by directly transforming $J$ using $\mathcal{T}$, because of the additional terms ($\mp v^{J}_{rs}$). When constructing the generalized Fock matrix $\mathcal{F}$, these additional terms, however, will be canceled out by adding $\mathcal{V}_1$.

$$
\begin{aligned}
\mathcal{K}^{\alpha}_{ps} &= \sum_{qr} \mathcal{D}^{\alpha}_{qr}\mathcal{V}^{\alpha\alpha}_{pqrs} \\
&= \sum_{qr} \mathcal{D}^{\alpha}_{qr} V_{pqrs} \\
&= \sum_{qr} \gamma^{\alpha}_{qr} V_{pqrs} \\
&= \left(\mathcal{T}(K^{\alpha})\right)_{ps},
\end{aligned}
\tag{S37}
$$

$$
\begin{aligned}
\mathcal{K}^{\beta}_{ps} &= \sum_{qr} \mathcal{D}^{\beta}_{qr}\mathcal{V}^{\beta\beta}_{pqrs} \\
&= \sum_{qr} \mathcal{D}^{\beta}_{qr} V_{pqrs} \\
&= \sum_{qr} -\gamma^{\beta}_{qr} V_{pqrs} + S^{-1}_{qr} V_{pqrs} \\
&= \left(\mathcal{T}(K^{\beta})\right)_{ps} + v^{K}_{ps}.
\end{aligned}
\tag{S38}
$$

Note that this result is different from that obtained by directly transforming $K$ using $\mathcal{T}$, because of the additional term ($v^{K}_{ps}$). This additional term is canceled out when adding with $\mathcal{V}_1$.

$$
\begin{aligned}
\mathcal{K}^{\alpha\beta}_{ps} &= \sum_{qr} \mathcal{D}^{\alpha\beta}_{qr}\mathcal{V}^{\alpha\beta}_{pqrs} \\
&= \sum_{qr} -\mathcal{D}^{\alpha\beta}_{qr} V_{pqrs} \\
&= \sum_{qr} -\kappa^{\alpha\beta}_{qr} V_{pqrs}.
\end{aligned}
\tag{S39}
$$

Note this term does not appear in the normal-state unrestricted Hartree-Fock (UHF) potential.

- Frozen-core approximation and DFT as the low-level theory

  The frozen-core potential ($\mathcal{J}^{\text{core}}$ and $\mathcal{K}^{\text{core}}$) and core energy ($\mathcal{E}^{\text{core}}$) can be calculated using Eq. (S33) and (S34) with $\mathcal{D}^{\text{core}}$. In a density functional theory (DFT) Kohn-Sham Hamiltonian, the $\mathcal{K}$ and $v^{K}$ matrices should be ignored for pure functionals, or be scaled with the hybrid parameter of hybrid functionals (e.g., $x^{\text{hyb}} = 1/4$ in the PBE0 functional).

### 1.1.3 DMET algorithms for superconducting states

In this section, we discuss how to perform a DMET calculation of a doped cuprate. The following algorithm is implemented in LIBDMET [62, 25].

1. Set up the cell and the lattice. The doping requires modifying the total number of electrons in the cell (within either the rigid-band or virtual crystal approximation).

2. Mean-field calculation of the doped system (e.g., using unrestricted orbitals and the PBE0 functional as in this work).

3. Local (core, valence, and virtual) orbitals are constructed, $C^{\text{AO,core}}$, $C^{\text{AO,non-core}}$ (see Sec. 1.2 for details).

4. Partial p-h transform: Construct $\mathcal{C}^{\text{AO,core}}$, $\mathcal{C}^{\text{AO,non-core}}$, $\mathcal{S}$, $\mathcal{H}_1$, $\mathcal{H}_0$, $\mathcal{V}_1$, $\mathcal{V}_0$, $\mathcal{D}^{\text{core}}$, $\mathcal{J}^{\text{core}}$, $\mathcal{K}^{\text{core}}$, $\mathcal{E}^{\text{core}}$.

   The resulting lattice Hamiltonian (all in a local orthogonal AO basis [62] without core orbitals, **k**-point labels are omitted for clarity) is

   $$\mathcal{H} = \mathcal{H}_1 + \mathcal{V}_1 + \mathcal{J}^{\text{core}} - x^{\text{hyb}}\mathcal{K}^{\text{core}} + \mathcal{E}, \tag{S40}$$

   where the constant term $\mathcal{E} = E^{\text{nuc}} + \mathcal{E}^{\text{core}} + \mathcal{H}_0 + \mathcal{V}_0$ includes the nuclear-nuclear repulsion energy, frozen-core energy, as well as the vacuum terms from the p-h transformation. Note that here $\mathcal{V}_1$ and $\mathcal{V}_0$ will also include a factor $x^{\text{hyb}}$ if DFT is used.

   The generalized Fock matrix comes from a direct transformation of the mean-field unrestricted DFT Kohn-Sham Fock matrices,

   $$\mathcal{F} = \begin{bmatrix} F^{\alpha} & 0 \\ 0 & -F^{\beta} \end{bmatrix}. \tag{S41}$$

   To constrain the expectation value of the physical particle number, an additional chemical potential is included in $\mathcal{H}$ and $\mathcal{F}$,

   $$\mathcal{M} = \begin{bmatrix} -\mu S & 0 \\ 0 & \mu S \end{bmatrix}. \tag{S42}$$

5. Set up the correlation potential.

   $$u^{\text{corr}} = \begin{bmatrix} 0 & \Delta^{\alpha\beta} \\ \Delta^{\alpha\beta\dagger} & 0 \end{bmatrix}. \tag{S43}$$

   The particle-number non-conserving part is from the correlation potential and is determined self-consistently through DMET. We also constrain the form of $\Delta$ to be a subset of local orbitals (the 3-band orbitals) and the point group symmetry ($C_{2h}$) is applied. The initial guess corresponds to a small $d$-wave potential among the four Cu $3d_{x^2-y^2}$ orbitals. In principle, we could also include the diagonal part in $u^{\text{corr}}$ to determine the magnetic order self-consistently, but to reduce the complexity of the problem, we do not consider this degree of freedom in this work.

6. DMET main loop.

   (a) Diagonalize the generalized lattice Fock matrix $\mathcal{F} + \mathcal{M} + u^{\text{corr}}$ to obtain the generalized density matrix $\mathcal{D}$ and determine the lattice chemical potential $\mu$ that ensures the correct electron number (Eq. (S16)). Note that a small-temperature smearing (e.g., $\beta = \frac{1}{k_{\text{B}}T} = 1000$) is used at finite doping because the lattice Fock matrix gap can be either small or vanishing.

   (b) Construct the embedding (impurity + bath) orbitals $\mathcal{C}$ from $\mathcal{D}$ using Eq. (S27) and Eq. (S28).

   (c) Construct the embedding Hamiltonian from the embedding orbitals $C$ and p-h transformed integrals[Eq. (S29) and (S31)]. Note that the DFT $v^{\text{xc}}$ does not enter into the embedding Hamiltonian, only the Hartree-Fock embedding-core interaction enters, so there is no double-counting in the many-body solution.

(d) Solve the embedding problem with generalized solvers (e.g., the coupled cluster method introduced in Sec. 1.3). We also need to determine the chemical potential of the embedding problem $\mu^{\mathrm{emb}}$ such that the physical electron number (Eq. (S16)) is the same as that obtained from the doped lattice (omitting the electron count from the core orbitals),

$$N^{\mathrm{elec,emb}} = N^{\mathrm{elec,latt\ without\ core}} \tag{S44}$$

Currently, $\mu^{\mathrm{emb}}$ is determined at the Hartree-Fock level. We find that the difference in the number of electrons between HF and CC (in the last DMET iteration after self-consistency) is about 0.15 $e$ per unit cell, which is 0.7 % of the valence electrons.

(e) Compute $\mathcal{D}^{\mathrm{emb}}$, transform it back to the original electron representation, and analyze the magnetic and pairing orders.

(f) Fit $u^{\mathrm{corr}}$ in the lattice problem by diagonalizing $\mathcal{F} + \mathcal{M} + u^{\mathrm{corr}}$ and using least-squares correlation potential fitting. We only fit the anomalous part of the density matrix subblocks (3-band orbital subblocks of $\mathcal{D}^{\alpha\beta}$).

(g) Extrapolate $\mathcal{D}$ and go back to (a) until the change $||\mathcal{D}_{i+1} - \mathcal{D}_i||$ is sufficiently small.

## 1.2 Localized orbitals for metallic systems

The ab initio DMET calculation is formulated in a local atomic orbital basis [62] constructed using the intrinsic atomic orbital (IAO) formalism [69]. The IAO formalism is usually formulated at zero-temperature, however, as discused above we use finite-temperature smearing in the lattice problem, thus we need to extend the intrinsic atomic orbital construction to this case. We also discuss how to define the core IAOs in the frozen-core calculations.

- k-adapted IAO [62]

  IAO can be viewed as a set of AO-character based projected Wannier functions. The key ingredients for IAO construction are the occupied MOs $\{|\psi_m\rangle\}$ and two sets of pre-defined localized bases, $B_1$ and $B_2$. $B_1$ is the normal AO basis used in the mean-field calculation (labeled by $\mu, \nu, \cdots$) and $B_2$ is the reference minimal basis set (labeled by $\rho, \sigma, \cdots$). $B_1$ usually contains the space of $B_2$ and the extra part reflects the polarization. The goal of IAO construction is to obtain a set of AO-like orbitals that contain the occupied space but have the size of the small basis set $B_2$. To achieve this, we first define the *depolarized* MOs $\{|\bar{\psi}_m\rangle\}$ by projecting the MOs to $B_2$, then back to $B_1$,

$$|\bar{\psi}_m\rangle = \mathrm{orth}\Big(P^{B_1} P^{B_2} |\psi_m\rangle\Big), \tag{S45}$$

  where $P$ is the resolution of identity (or projector) of AOs,

$$P^{B_1}_{\mu\nu} = \sum_{\mu\nu} |\phi_\mu\rangle S^{B_1}_{\mu\nu} \langle\phi_\nu| \,. \tag{S46}$$

  Using the depolarized MO projector $\bar{O} \equiv \sum_m |\bar{\psi}_m\rangle\langle\bar{\psi}_m|$, we can split the $B_2$ set into occupied ($\bar{O}\,|\phi_\rho\rangle$) and virtual spaces $(1 - \bar{O})\,|\phi_\rho\rangle$. The IAOs $\{|w_i\rangle\}$ are obtained by further projecting these two subspace bases onto their polarized counterparts ($O \equiv \sum_m |\psi_m\rangle\langle\psi_m|$ and $1 - O$) and applying Löwdin orthogonalization,

$$|w_i\rangle = \mathrm{orth}\big\{\big[O\bar{O} + (1 - O)(1 - \bar{O})\big]\,|\phi_\rho\rangle\big\}. \tag{S47}$$

  In periodic systems, the quantities in the above equations should be understood to carry $\mathbf{k}$ labels, e.g. $|\phi_\mu\rangle \to |\phi^{\mathbf{k}}_\mu\rangle$ is a crystal AO, and $S^{B_1} \to S^{\mathbf{k},B_1}$ is the corresponding overlap matrix. These quantities are already evaluated in the mean-field calculations. The only thing we need additionally is the overlap matrix $S_{12}$ between basis $B_1$ and $B_2$, which can be evaluated directly,

$$S^{\mathbf{k},B_1,B_2}_{\mu\rho} = \sum_{\mathbf{T}} \int \mathrm{d}\mathbf{r} e^{\mathrm{i}\mathbf{k}\cdot\mathbf{T}} \phi^*_\mu(\mathbf{r}) \phi_\rho(\mathbf{r} - \mathbf{T}), \tag{S48}$$

  where the summation is over the periodic images $\mathbf{T}$. After the IAOs are constructed, the $\mathbf{k}$-adapted PAOs are obtained by projecting out the IAO components from the AOs at each $\mathbf{k}$-point.

- IAO with frozen-core orbitals

  When dealing with frozen-core calculations, the construction of IAOs needs to be adjusted. The key idea is that $B_1$ basis now becomes a subset of MOs (i.e., only treat the non-core part of MOs), $C_{pP}$ (we use capital letters $P, Q, \cdots$ for non-core indices). So the new occupied coefficients in the non-core basis become

$$C^{\text{occ}}_{pm} \to C^{\text{occ}}_{Pm} = \sum_{pq} C^{\dagger}_{Pq} S_{qp} C^{\text{occ}}_{pm}. \tag{S49}$$

  The large basis overlap $S_1$ matrix changes as

$$S_{pq} \to S_{PQ} = \sum_{pq} C^{\dagger}_{Pp} S_{pq} C_{qQ}. \tag{S50}$$

  Similarly, the inter-molecular overlap $S_{12}$ reads,

$$S_{p\mu} \to S_{P\mu} = \sum_{p} C^{\dagger}_{Pp} S_{p\mu}. \tag{S51}$$

  The core and non-core IAOs are then constructed separately.

- IAO with finite-temperature smearing

  For the large basis $O$, we can replace the occupied projector by a finite-temperature projector

$$O = \sum_{m}^{\text{occ}} |\psi_m\rangle \langle\psi_m| \to \sum_{m} |\psi_m\rangle f_m \langle\psi_m| , \tag{S52}$$

  where $f_m = \frac{1}{\mathrm{e}^{\beta(\varepsilon_m - \mu)}+1}$. The depolarized MO projector $\bar{O}$ becomes,

$$\bar{O} = \sum_{m}^{\text{occ}} |\bar{\psi}_m\rangle\langle\bar{\psi}_m| \to \sum_{\varepsilon_m < \mu} |\bar{\psi}_m\rangle\langle\bar{\psi}_m| . \tag{S53}$$

  We cannot use the same formula in Eq. (S52) since $\{\bar{\psi}\}$ is not an orthogonal orbital set. Instead, we use the depolarized MOs below the Fermi level $\mu$. This definition makes the IAOs still approximately span the "occupied space" of the smeared wavefunction. The deviation of the electron number trace in this IAO space (reflecting the inexact span of the occupied orbitals) is of order $1/\beta$.

### 1.3 Ab initio many-body impurity solver

As discussed above and in the main text, within the Nambu formulation, the DMET quantum many-body problem corresponds to the quantum chemistry problem of determining the ground-state of a Hamiltonian in the generalized spin-orbital language. The corresponding quantum chemistry solvers are then called "generalized" solvers. Below we describe three solvers used in this work: generalized coupled cluster with singles and doubles; generalized tailored coupled cluster with singles and doubles, and generalized ab initio density matrix renormalization group. Results from the tailored CC and ab initio DMRG solvers can be found in Sec. 3.1.

- Generalized coupled cluster

  The main solver we use is the generalized coupled cluster singles and doubles (GCCSD) solver [70], which can be readily applied to *ab initio* embedding Hamiltonians with hundreds of orbitals. It is based on a wavefunction ansatz of the form

$$|\Psi\rangle = \mathrm{e}^{\hat{T}_1 + \hat{T}_2} |\Phi\rangle , \tag{S54}$$

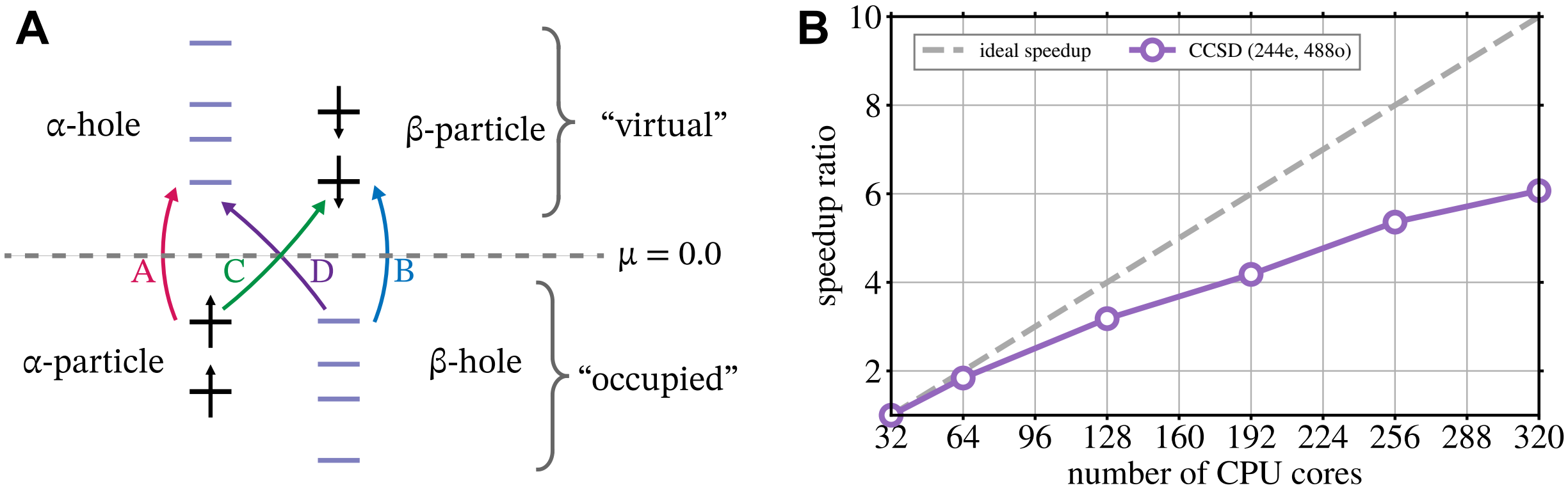

Figure S1: (A) Illustration of excitations in a superconducting state in the Nambu representation: The orbital space is split into occupied and virtual spaces. The 'occupied' orbitals arise from the mixing of $\alpha$ particles and $\beta$ hole states; while the 'virtual' orbitals arise from the mixing of $\alpha$ holes and $\beta$ particle states. The chemical potential $\mu = 0$. Different types of one-body excitations: A and B types of excitations are the normal-state particle-hole excitations, whereas C and D are superconducting types of excitation: C and D label the particle-particle and hole-hole channels respectively. (B) MPI parallel efficiency of generalized CCSD calculation in a half-filled cuprate impurity problem (488 spin orbitals).

where $|\Phi\rangle$ is a reference Slater determinant (a generalized HF determinant with a constrained average electron number) and the cluster excitation operators are defined as,

$$
\begin{aligned}
\hat{T}_1 &= \sum_{ia} t_i^a c_a^\dagger c_i, \\
\hat{T}_2 &= \frac{1}{4} \sum_{ijab} t_{ij}^{ab} c_a^\dagger c_b^\dagger c_j c_i.
\end{aligned}
\tag{S55}
$$

Since our orbitals are mixtures of particles and holes, the corresponding excitations contain both normal-state particle-hole type excitations, as well as superconducting type particle-particle (or hole-hole) excitations (as illustrated in Fig. S1A). The 'occupied' orbitals arise from the mixing of $\alpha$ particles and $\beta$ hole states; while the 'virtual' orbitals arise from the mixing of $\alpha$ holes and $\beta$ particle states. The whole spectrum is symmetric with respect to the chemical potential $\mu = 0$. In the limit of a non-superconducting state (pairing field $\Delta^{\alpha\beta} = 0$), the $\alpha$-particle and $\beta$-holes are not mixed and the generalized Slater determinant calculation can be mapped onto an unrestricted Slater determinant.

In gapless or small gap problems, there can be difficulties converging the CCSD equations. The CCSD amplitudes are determined from the residual equations [29],

$$
\begin{aligned}
r_i^a &= \left\langle \Phi_i^a \middle| \mathrm{e}^{-\hat{T}_1 - \hat{T}_2} H \mathrm{e}^{\hat{T}_1 + \hat{T}_2} \middle| \Phi \right\rangle = 0, \\
r_{ij}^{ab} &= \left\langle \Phi_{ij}^{ab} \middle| \mathrm{e}^{-\hat{T}_1 - \hat{T}_2} H \mathrm{e}^{\hat{T}_1 + \hat{T}_2} \middle| \Phi \right\rangle = 0.
\end{aligned}
\tag{S56}
$$

Solving these equations is a non-linear root-finding problem and the standard method is to use Jacobian iteration with direct inversion of the iterative space (DIIS) [71]. This uses a preconditioner of the form

$$
\mathcal{P} = \frac{1}{\varepsilon_i - \varepsilon_a} \propto \frac{1}{\text{gap}}
\tag{S57}
$$

which does not work for small gap problems.

Various strategies can be employed to improve convergence, such as a regularized preconditioner,

$$
\mathcal{P} = \frac{1}{\varepsilon_i - \varepsilon_a + \delta} \quad \text{or} \quad \frac{1}{\varepsilon_i - \varepsilon_a} (1 - \mathrm{e}^{-\kappa(\varepsilon_i - \varepsilon_a)}).
\tag{S58}
$$

We instead use a Newton-Krylov method, which approximates the inverse of the Jacobian $\mathbf{J}^{-1}$ in a Krylov subspace and solves for the root in an inexact Newton way (we use a level-shift form of the preconditioner here as in Eq. (S58) with $\delta = 0.01$ a.u.). For the $k^{\text{th}}$ iteration, the CCSD amplitudes $\mathbf{t}^k$ are updated as,

$$\mathbf{t}^{k+1} = \mathbf{t}^k - \mathbf{J}^{-1}\,\mathbf{r}\left[\mathbf{t}^k\right]. \tag{S59}$$

We have found this scheme to be more reliable for systems with near-orbital degeneracy than the traditional DIIS method [72, 73]. We implemented the Newton-Krylov GCCSD in the MPI4PySCF package with efficient MPI parallelism [74], see Fig. S1B for a cuprate impurity problem (488o, 244e) scaled up to 320 CPU cores.

As the CCSD method is inherently approximate in nature, it is important to evaluate its accuracy. To accomplish this, we benchmark the CCSD against the tailored coupled cluster method and the *ab initio* DMRG approach.

- Tailored coupled cluster

  Tailored coupled cluster theory (TCC) was introduced as a method to treat systems with stronger correlation than can usually be treated by standard coupled cluster methods [75]. The basic idea is to treat part of the correlation by an exact method (full configuration interaction/exact diagonalization) and to treat the remaining correlation by the coupled cluster ansatz. The TCC wavefunction separates the cluster operators by categorizing them as either acting on the active ($T_{\text{act}}$) or external ($T_{\text{ext}}$) space of orbitals,

$$|\Psi\rangle = \mathrm{e}^{\hat{T}_{\text{act}}+\hat{T}_{\text{ext}}}|\Phi\rangle \tag{S60}$$

  where the active space orbitals are the most strongly correlated ones. Specifically, in TCCSD theory, both $\hat{T}_{\text{act}}$ and $\hat{T}_{\text{ext}}$ are limited to single and doubles excitations,

$$\hat{T}_{\text{act}} = \hat{T}_{\text{act},1} + \hat{T}_{\text{act},2} \quad \hat{T}_{\text{ext}} = \hat{T}_{\text{ext},1} + \hat{T}_{\text{ext},2} \tag{S61}$$

  The active space component of the amplitudes is calculated using the full configuration interaction coefficients within the active space, corresponding to single and double excitations, $c_i^a$, $c_{ij}^{ab}$. Notably, these amplitudes are fixed in the calculation.

$$(t_{\text{act}})_i^a = \frac{c_i^a}{c_0}, \quad (t_{\text{act}})_{ij}^{ab} = \frac{c_{ij}^{ab}}{c_0} - \frac{c_i^a c_j^b - c_i^b c_j^a}{c_0^2} \quad a, b, i, j \in \text{act} \tag{S62}$$

  Coefficients corresponding to higher excitations within the active space are disregarded. The external component of the amplitudes is then determined by solving the amplitude equations consistent with Eq. (S56). The density matrix for the TCCSD method is computed in this study using the Lagrangian approach.

- Density matrix renormalization group

  The ab initio density matrix renormalization group (DMRG) [76, 77] is regarded as a nearly exact method for quantum chemistry problems with several tens of orbitals. In this work, we use it as a benchmark tool to obtain an accurate solution of the impurity problem within a reduced active space of orbitals determined by the CC natural orbitals (the eigenvectors of CC density matrix) [78, 79].

### 1.4 Correlation potential matching between lattice and embedded impurity

In the self-consistent DMET cycle, we carry out self-consistency with respect to the superconducting order. We restrict the self-consistency to the (three-band block of) the anomalous part of the density matrix $\mathcal{D}^{\alpha\beta}$ of the lattice and embedding problems (although the pairing is non-zero outside of this block). We match the lattice and impurity orders using a least-squares fitting procedure [80], with an additional chemical potential to exactly match the electron number. In other words, after the embedding density matrix $\mathcal{D}^{\text{emb}}$ is computed, we perform the minimization

$$\min_{u^{\text{corr}}} ||\mathcal{D}_{ij}^{\alpha\beta,\mathbf{R}=\mathbf{0}}[u^{\text{corr}}] - \mathcal{D}_{ij}^{\alpha\beta,\text{emb}}|| \tag{S63}$$

where $ij$ are limited to the three-band orbitals indices, and $u^{\mathrm{corr}}$ is the correlation potential on the three-band orbitals that obey $C_{2h}$ point-group symmetry. $\mathcal{D}^{\mathbf{R}=\mathbf{0}}$ is obtained from the lattice Fermi-Dirac function,

$$\left(\mathcal{F}^{\mathbf{k}} + \mathcal{M} + u^{\mathrm{corr}}\right)\mathcal{C}^{\mathbf{k}} = \mathcal{C}^{\mathbf{k}}\mathcal{E}^{\mathbf{k}} \tag{S64}$$

$$\mathcal{D}^{\mathbf{k}} = \mathcal{C}^{\mathbf{k}} \frac{1}{\exp\left(\mathcal{E}^{\mathbf{k}}/k_{\mathrm{B}}T\right) + 1} \mathcal{C}^{\mathbf{k}\dagger} \tag{S65}$$

The chemical potential $\mu$ in $\mathcal{M}$ is additionally fitted during the minimization procedure to ensure the lattice problem always has the correct expectation value of the electron number (Eq. (S16)). To ensure the stability of this procedure, a small smearing temperature $T$ is applied in lattice problem. The default temperature we use in this work is $\beta = 1000$. Additional results with $\beta = 2000$ and 5000 are given in Sec. 3.3.

### 1.5 Analysis methods

- Order parameters

  In our work on the parent state [25], we defined charge, magnetic, and bond orders. We collect the relevant formulae here. The charge of local orbital $i$ is,

$$n_i = \gamma_{ii}^{\alpha} + \gamma_{ii}^{\beta}, \tag{S66}$$

  and the local magnetic moment of orbital $i$ is defined as the charge difference between up ($\alpha$) and down ($\beta$) spin densities,

$$m_i = \gamma_{ii}^{\alpha} - \gamma_{ii}^{\beta}. \tag{S67}$$

  The 2-center Mayer bond order [81] between atoms $A$ and $B$ (or two subsets of orbitals) is defined as,

$$b_{AB} = \sum_{\sigma} b_{AB}^{\sigma} = 2 \sum_{\sigma} \sum_{i \in A} \sum_{j \in B} \left(\gamma^{\sigma} S\right)_{ji} \left(\gamma^{\sigma} S\right)_{ij}, \tag{S68}$$

  where $\gamma^{\sigma}$ is the one-particle density matrix with spin $\sigma$ and $S$ is the overlap matrix of the local basis.

  To characterize the SC order, we define real-space and orbital-resolved SC order parameters.

  In one-band models and three-band Hubbard models, the SC order parameter is typically evaluated as the average of the Cu-Cu pairing components,

$$\begin{aligned} m_{\mathrm{SC}} &= \sum_{\langle ij \rangle} \frac{1}{\sqrt{2}} \eta_{ij}^{\mathrm{SC}} \left(\langle d_{i\alpha} d_{j\beta} \rangle + \langle d_{j\alpha} d_{i\beta} \rangle\right) \\ &= \sum_{\langle ij \rangle} \frac{1}{\sqrt{2}} \eta_{ij}^{\mathrm{SC}} \left(\kappa_{ij}^{\alpha\beta} + \kappa_{ij}^{\alpha\beta\dagger}\right), \end{aligned} \tag{S69}$$

  where $d$ denotes fermion operators on the Cu sites, and $\langle \cdots \rangle$ limits the summation such that only the pairing between nearest Cu sites is taken into account. $i, j$ are local orbital indices, $\alpha$ and $\beta$ labels the two spin channels (i.e, the SC order is computed from the $\alpha$-$\beta$ coupling block of density matrix). The $d$-wave superconducting structure factor $\eta^{\mathrm{SC}}$ is defined as

$$\eta_{ii'}^{\mathrm{SC}} = \begin{cases} +1, \text{ if } \mathbf{R}_i - \mathbf{R}_{i'} = \pm \mathbf{e}_x, \\ -1, \text{ if } \mathbf{R}_i - \mathbf{R}_{i'} = \pm \mathbf{e}_y. \end{cases} \tag{S70}$$

  where $\mathbf{e}_{x(y)}$ is the unit-cell lattice vector in the $x(y)$ direction. For $s$-wave Cu pairing, we use $\eta = 1$ for all terms.

  In an *ab initio* calculation, it is interesting to explore the multi-orbital nature of the pairing. The above analysis can be extended to individual orbital-orbital pairing components by a suitable modification of the definition of the lattice vectors. However, it is useful to define a summarized version of the total pairing. Note that there is no unique way to define this (as there are multiple ways to reduce to a single scalar). In this work, we define the total atomic pairing order, given a set of $\{\phi\}$ attached to each atom, as

$$\begin{aligned} m_{\mathrm{SC}} &= \sum_{ij} \left| \sum_{\langle IJ \rangle} \frac{1}{\sqrt{2}} \eta_{IJ}^{\mathrm{SC}} \left( \left\langle \phi_{i\alpha}^{I} \phi_{j\beta}^{J} \right\rangle + \left\langle \phi_{j\alpha}^{J} \phi_{i\beta}^{I} \right\rangle \right) \right| \\ &= \sum_{ij} \left| \sum_{\langle IJ \rangle} \frac{1}{\sqrt{2}} \eta_{IJ}^{\mathrm{SC}} \left( \kappa_{ij}^{\alpha\beta IJ} + \kappa_{ij}^{\beta\alpha IJ} \right) \right|, \quad \text{[used in the main text]} \end{aligned} \tag{S71}$$

where $i$ $(j)$ loops all of the local orbitals of atom $I$ $(J)$, and the phase factors $\eta$ are the same as before. (Note that the intracell order is not included in the summation).

Besides Eq. (S71) (the main form of SC order we used in this work), we examine two other definitions of pairing order. One is to resolve the total pairing coupling of a given symmetry between unit cells. For example, in the $2 \times 2$ impurity, we can compute the angular momentum resolved orders as,

$$\begin{aligned} m^{s,\mathrm{intra}} &= \sum_{\mathbf{R}=\mathbf{0},\mathbf{1},\mathbf{2},\mathbf{3}} \sum_{ij} \left| \frac{1}{\sqrt{2}} \left( \kappa_{ij}^{\mathbf{RR}} + \mathrm{H.c} \right) \right| \\ m^{s} &= \sum_{ij} \left| \frac{1}{\sqrt{2}} \left[ \frac{1}{2} \left( \kappa_{ij}^{\mathbf{01}} + \kappa_{ij}^{\mathbf{12}} + \kappa_{ij}^{\mathbf{23}} + \kappa_{ij}^{\mathbf{30}} \right) + \mathrm{H.c} \right] \right| \\ m^{p_x} &= \sum_{ij} \left| \frac{1}{\sqrt{2}} \left[ \frac{1}{\sqrt{2}} \left( \kappa_{ij}^{\mathbf{03}} - \kappa_{ij}^{\mathbf{30}} \right) + \mathrm{H.c} \right] \right| \\ m^{d} &= \sum_{ij} \left| \frac{1}{\sqrt{2}} \left[ \frac{1}{2} \left( \kappa_{ij}^{\mathbf{01}} - \kappa_{ij}^{\mathbf{12}} + \kappa_{ij}^{\mathbf{23}} - \kappa_{ij}^{\mathbf{30}} \right) + \mathrm{H.c} \right] \right|, \end{aligned} \tag{S72}$$

where $\mathbf{0}, \mathbf{1}, \mathbf{2}, \mathbf{3}$ are labels of unit cells in a $2 \times 2$ impurity. We show the angular momentum resolved components of the total order in Fig. S4B (see als for the cell arrangement).

We also study the real-space amplitude of the anomalous density matrix on the lattice

$$\kappa^{\alpha\beta}(\mathbf{R}_0, \mathbf{r}) = \sum_{ij} \phi_i^{\alpha}(\mathbf{R}_0) \kappa_{ij}^{\alpha\beta} \phi_j^{\beta *}(\mathbf{r}). \tag{S73}$$

where $\mathbf{R}_0$ is chosen as a reference point near a Cu atom. This is visualized in Fig. 2C of the main text.

- Coupled cluster amplitudes decomposition

  To understand the fluctuations driving pairing, we perform various analyses of the CCSD amplitude tensors $t_i^a$ and $t_{ij}^{ab}$, where $ij$ $(ab)$ are occupied (virtual) orbital labels.

  The first analysis is to check the correlation between the SC order and different blocks of CCSD amplitudes. As we have shown in Fig. S1A, the MOs are mixture of holes and particles and there is an arbitrary unitary rotation among occupied (or virtual) space. In order to label each MO with its largest character, we need to localize the occupied and virtual orbitals separately. The localization then re-mixes the orbital characters and gives a maximal separation between the $\alpha$ and $\beta$ "spin" blocks. After the localization (by selected columns of density matrix, SCDM [82]), we can then attach each MO with a spin label. This analysis is valid for states where the particle-number-breaking sector is not large and the particle/hole characters can be largely disentangled.

  After the localization procedure, the CCSD amplitudes then have block-wise structure. For instance, the $T_1$ amplitudes can be split into four blocks, as shown in Table S1. The first two blocks conserve the particle number while the third and fourth blocks change the particle number by 2, reflecting the superconducting pairing type excitations. Similarly, $T_2$ can be classified as normal and superconducting amplitudes based on the change in particle number. The maximal change is 4. One can see that the main excitation is the normal-state particle-hole excitations, in particular the amplitude $t_{\alpha\beta}^{\alpha\beta}$ which includes the AFM type spin interactions. The main contribution in the superconducting blocks are of $N \pm 2$ particle number sector and $N \pm 4$ type excitations can be ignored.

Table S1: Decomposition of CCSD amplitudes based on the orbital character. The Frobenius norm values are taken from the calculation of CCO at normal pressure.

| Amplitudes | $i$ | $a$ | $j$ | $b$ | particle number | norm |
|---|---|---|---|---|---|---|
| $t_{\alpha}^{\alpha}$ | $\alpha$ (p) | $\alpha$ (h) | | | $N$ | 0.495 |
| $t_{\beta}^{\beta}$ | $\beta$ (h) | $\beta$ (p) | | | $N$ | 0.495 |
| $t_{\alpha}^{\beta}$ | $\alpha$ (p) | $\beta$ (p) | | | $N-2$ | 0.084 |
| $t_{\beta}^{\alpha}$ | $\beta$ (h) | $\alpha$ (h) | | | $N+2$ | 0.079 |
| $t_{\alpha\alpha}^{\alpha\alpha}$ | $\alpha$ (p) | $\alpha$ (h) | $\alpha$ (p) | $\alpha$ (h) | $N$ | 0.539 |
| $t_{\alpha\beta}^{\alpha\beta}$ | $\alpha$ (p) | $\alpha$ (h) | $\beta$ (h) | $\beta$ (p) | $N$ | 0.663 |
| $t_{\alpha\beta}^{\beta\beta}$ | $\alpha$ (p) | $\beta$ (p) | $\beta$ (h) | $\beta$ (p) | $N-2$ | 0.065 |
| $t_{\alpha\alpha}^{\beta\beta}$ | $\alpha$ (p) | $\beta$ (p) | $\alpha$ (p) | $\beta$ (p) | $N-4$ | 0.004 |
| $\cdots$ | $\cdots$ | $\cdots$ | $\cdots$ | $\cdots$ | $\cdots$ | $\cdots$ |

The second analysis is the SVD decomposition of $T_2$ amplitudes. To explain our procedure, it is useful to establish the correspondence between $T_2$ amplitudes in the Nambu representation and without the Nambu transformation, thus we first assume that there is no pairing field. Then the initial generalized Hartree-Fock ground-state in the Nambu representation is the same as an unrestricted Hartree-Fock ground-state of the original electrons. Using the Nambu creation and annihilation basis $(c^\dagger, c)$, the $\hat{T}_2$ operator is

$$\hat{T}_2 = \frac{1}{4} \sum_{ijab} t_{ij}^{ab} c_a^\dagger c_b^\dagger c_j c_i \tag{S74}$$

where $i, j$ are the Nambu "occupied" indices, and $a, b$ are the Nambu "virtual" indices (see Fig. S1A). Through the particle-hole transformation, we see that $c_i \in \{a_{I\alpha}, a_{A\beta}^\dagger\}$ and $c_a^\dagger \in \{a_{A\alpha}^\dagger, a_{J\beta}\}$, where $I, J$ are the electron occupied orbital indices, and $A, B$ are electron virtual indices. Then, for pairs of fermionic operators, we have relations like

$$\begin{aligned} c_i c_a^\dagger &\in \{a_{I\alpha} a_{A\alpha}^\dagger, a_{I\alpha} a_{J\beta}, a_{A\beta}^\dagger a_{A\alpha}^\dagger, a_{A\beta}^\dagger a_{J\beta}\} \\ c_i c_j &\in \{a_{I\alpha} a_{J\alpha}, a_{I\alpha} a_{A\beta}^\dagger, a_{A\beta}^\dagger a_{B\beta}^\dagger, a_{A\beta}^\dagger a_{I\alpha}\} \\ &\cdots \end{aligned} \tag{S75}$$

where the first pair operator $c_i c_a^\dagger$ contains 4 types of single excitations (see also Fig. S1A and the $t_1$ amplitudes in Table S1), where all of them conserve spin ($\Delta S_z = 0$), whereas the second pair $c_i c_j$ changes spin ($\Delta S_z = -1$). If there is a pairing field (so that the generalized Hartree-Fock ground-state is no longer mappable onto an unrestricted Hartree-Fock ground-state) then $c_i$ and $c_a^\dagger$ become linear combinations of their electron particle and electron hole components, but this does not change the spin nature of the pair operators $c_i c_j$, $c_i c_a^\dagger$.

Thus, we can consider two different SVD decompositions of $T_2$ from grouping the indices in two ways:

$$\begin{aligned} t_{ia,jb} &= \sum_s U_{ia,s}^0 \Sigma_s^0 V_{s,jb}^{0\dagger} \text{ (spin conserving),} \\ t_{ij,ab} &= \sum_s U_{ij,s}^{\pm 1} \Sigma_s^{\pm 1} V_{s,ab}^{\pm 1\dagger} \text{ (spin fluctuating).} \end{aligned} \tag{S76}$$

Using these singular values, we rewrite $\hat{T}_2$,

$$\hat{T}_2 = \sum_s \Sigma_s^0 \hat{O}_s^0 \hat{\bar{O}}_s^0 \text{ (spin conserving),} \tag{S77}$$

$$\hat{T}_2 = \sum_s \Sigma_s^{\pm 1} \hat{O}_s^{\pm 1} \hat{\bar{O}}_s^{\pm 1} \text{ (spin fluctuating).} \tag{S78}$$

We can then check to see if the spectrum of singular values of $\Sigma$ is dominated by a few modes in the spin-conserving or spin-fluctuating decomposition, and we can also analyze the dominant singular vectors to see their effects on the charge and magnetic density. This type of fluctuation analysis is analogous to that performed for two-particle Green's functions e.g. in [58]. Alternatively, some related intuition is provided by a ground-state random phase approximation (RPA) analysis. Regarding the $\hat{O}$ above as (quasi)bosonic operators, we see that in the RPA, the $\hat{T}_2$ CC wavefunction can be written as

$$|\Psi\rangle = \mathrm{e}^{\frac{1}{2}\sum_{\mu\nu} t_{\mu\nu} b_\mu^\dagger b_\nu^\dagger} |0_\mathrm{B}\rangle \tag{S79}$$

where $b$ is a bosonic operator (defined on the boson vacuum $|0_\mathrm{B}\rangle$). Then, the above effects a canonical transformation, where the rotated particles (from the $T$ amplitude rotation) are

$$\tilde{b}_\mu = \sum_\nu \left( [1 - t^2]^{1/2}_{\mu\nu} b_\nu + ([1 - t^2]^{1/2} t)_{\mu\nu} b_\nu^\dagger \right) \approx b_\mu + \sum_\nu t_{\mu\nu} b_\nu^\dagger \tag{S80}$$

A dominant singular value in $t_{\mu\nu}$ thus defines a main bosonic degree of freedom that is "mixed in" to the original degrees of freedom.

# 2 Computational details

## 2.1 Atomic modelling of doping

The hole or electron doping of cuprates is usually generated by chemical doping, which involves substituting buffer layer ions or introducing additional ions. For example, $La_2CuO_4$ is typically doped by replacing some of the La (III) ions with Sr (II) ions; because the entire crystal remains neutral, this means that charge must be taken out of the cuprate plane, effectively doping the plane with holes. Another example particularly relevant to this work is the hole doping of Hg-based cuprates, where oxygen is introduced into the Hg buffer layer, whose charge must then be compensated by other ions, including from the $CuO_2$ plane. Note that the effective doping of the $CuO_2$ plane is not always simple to determine from the dopant atom concentration alone, as charge can be redistributed across many different atoms. Consequently, the best way to simulate chemical doping is to use explicit dopant atoms. Unfortunately, the corresponding supercells become large and thus difficult to treat.

A simple way to dope without using explicit dopants is provided by the *rigid band approximation* (RBA). In this case, one directly removes or adds charge to the system, and assumes that there is a neutralizing background charge. This is similar in spirit to doping in lattice models, which is performed simply by modifying the number of particles.

Another method is the *virtual crystal approximation* (VCA) [83, 32], where the nuclear potential is obtained by mixing different site compositions, e.g. if a lattice site has $x$ probability being occupied by atom A, and $(1 - x)$ probability by atom B, the VCA nuclear potential is defined as,

$$\bar{H}^{\mathrm{VCA}}_{\mathrm{nuc}}(\mathbf{r}) = x V^{\mathrm{A}}_{\mathrm{nuc}}(\mathbf{r}) + (1 - x) V^{\mathrm{B}}_{\mathrm{nuc}}(\mathbf{r}). \tag{S81}$$

This approximation is still very crude as one can imagine that the effect of a half-occupied oxygen ($Z = 8$) site is fundamentally different to the potential from a Be atom ($Z = 4$).

In the current work, we mainly use the RBA treatment of doping with VCA applied in a few cases as an additional benchmark. Results from the VCA treatment of CCO at different pressures are presented in Sec. 3.2.

We enforce a neutral total lattice and we consider crystals where the total number of electrons (summed over all sampled $\mathbf{k}$ points) is an even integer (the number of electrons per cell can obviously be a fractional number). We use a $2 \times 2$ supercell to allow superconducting orders. If we define the doping concentration as the number of additional charges per Cu, then this allows a minimal doping of $1/(4n_{\mathbf{k}})$, where 4 is the number of Cu per cell. For the VCA calculations on CCO, the VCA potential mixing was carried out at the Ca ion, whose nuclear charge was modified according to the doping.

## 2.2 Optimization of Gaussian basis sets

To enable compact and numerically well-conditioned calculations on the materials in this work, we optimized a new set of Gaussian bases. The structure of the Gaussian basis sets for Ca, Cu, Ba, La, and Hg used in this work is summarized in Table S2. These basis sets are of double-zeta quality and generated for the GTH pseudopotentials [84, 85] optimized for HF calculations [86] by following a protocol modified from Ref. [87], which is outlined below. The Gaussian basis sets for all other elements are taken from Ref. [87].

1. A series of uncontracted valence basis sets of different sizes is generated by minimizing the HF ground-state energy of a free atom. For Ca and Cu, where the $[4p]$ shell is unoccupied in the HF ground state, we use the ground-state CCSD correlation energy of a free atom. A similar treatment is applied for the $[6p]$ shell of Ba, La, and Hg.

2. For a given uncontracted valence basis, a series of polarization basis sets of different sizes is generated by minimizing the ground-state CCSD correlation energy of a free atom.

3. A series of full uncontracted basis sets (valence + polarization) is generated by combining the valence basis sets and the polarization basis sets obtained in the previous steps. The full uncontracted basis sets whose overlap matrix condition number evaluated for the reference solids listed in Table S3 is less than $10^{10}$ are kept.

4. For the full uncontracted basis sets surviving the previous step, the one that minimizes the average error of the atomization energy of the reference molecules listed in Table S3 is selected (the error is calculated against a large basis set discarded in the previous step.)

5. The selected full uncontracted basis set is contracted using the coefficients of the spherically averaged CCSD atomic natural orbitals [88]. Different contraction patterns have been tested, and the one that minimizes the average error of the molecular atomization energy is selected to make the final contracted basis set.

Table S2: Summary of the optimized Gaussian basis sets used in this work. Valence configuration refers to the electrons not frozen in the chosen GTH pseudopotentials. The structure of both the uncontracted (denoted by parenthesis) and contracted (denoted by bracket) basis sets are listed for both the valence basis and the polarization basis. The structure of the final contracted basis set is also included.

| Element | Valence configuration | Valence basis | Polarization basis | Final contracted basis |
|---|---|---|---|---|
| Ca | $[3s^2 3p^6 4s^2]$ | $(5s, 5p) \rightarrow [3s, 3p]$ | $(5d) \rightarrow [2d]$ | $[3s, 3p, 2d]$ |
| Cu | $[3s^2 3p^6 3d^{10} 4s^1]$ | $(5s, 6p, 4d) \rightarrow [3s, 3p, 2d]$ | $(2f) \rightarrow [1f]$ | $[3s, 3p, 2d, 1f]$ |
| Ba | $[5s^2 5p^6 6s^2]$ | $(5s, 6p) \rightarrow [3s, 3p]$ | $(6d) \rightarrow [2d]$ | $[3s, 3p, 2d]$ |
| La | $[5s^2 5p^6 5d^1 6s^2]$ | $(6s, 6p, 3d) \rightarrow [4s, 3p, 2d]$ | $(4f) \rightarrow [1f]$ | $[4s, 3p, 2d, 1f]$ |
| Hg | $[5d^{10} 6s^2]$ | $(4s, 4p, 5d) \rightarrow [3s, 3p, 3d]$ | $(2f) \rightarrow [1f]$ | $[3s, 3p, 3d, 1f]$ |

We tested different bases on CCO using unrestricted PBE0. From Fig. S2, it is clear that all the bases beyond the single-$\zeta$ level provide reasonable Birch-Murnaghan (B-M) equations of state (EOS) [89, 90]. The single-$\zeta$ basis (GTH-SZV), however, is too small to give the correct EOS.

In terms of the magnetic properties, the magnetic moment decreases when pressure increases (volume decreases) and the magnetic coupling $J$ increases. In the single-$\zeta$ basis the first trend is reversed, although the $J$ trend is qualitatively correct. The GTH-DZVP basis has a qualitatively correct trend but the absolute value of $J$ (141 meV) is still far away from the plane-wave reference number (217 meV). The new optimized GTH-cc-pVDZ basis and the all-electron triple-$\zeta$-level def2-TZVP basis are both in good agreement with the plane-wave reference (error in $\Delta J < 5$ meV). This confirms that the double-$\zeta$ level GTH-cc-pVDZ basis in this work provides a good balance between accuracy and efficiency.

Table S3: Reference solids and molecules used for determining the size and contraction pattern of the basis sets developed in this work.

| Element | Reference solids | Reference molecules |
|---|---|---|
| Ca | CaO, $CaF_2$ | CaH, $CaH_2$, CaF, $CaF_2$, CaO, CaS |
| Cu | Cu (fcc), CuF, CuO | CuH, $CuH_2$, $CuH_3$, CuF, $CuF_2$, CuO, CuOF, CuS |
| Ba | BaO, $BaF_2$ | BaH, $BaH_2$, BaF, $BaF_2$, BaO, BaS |
| La | LaN, La (fcc/hcp), $La_2O_3$ | LaH, $LaH_2$, $LaH_3$, LaF, $LaF_2$, $LaF_3$, LaO, LaN, LaS |
| Hg | HgO, Hg (fcc), HgS | HgH, $HgH_2$, HgF, $HgF_2$, HgO, HgS |

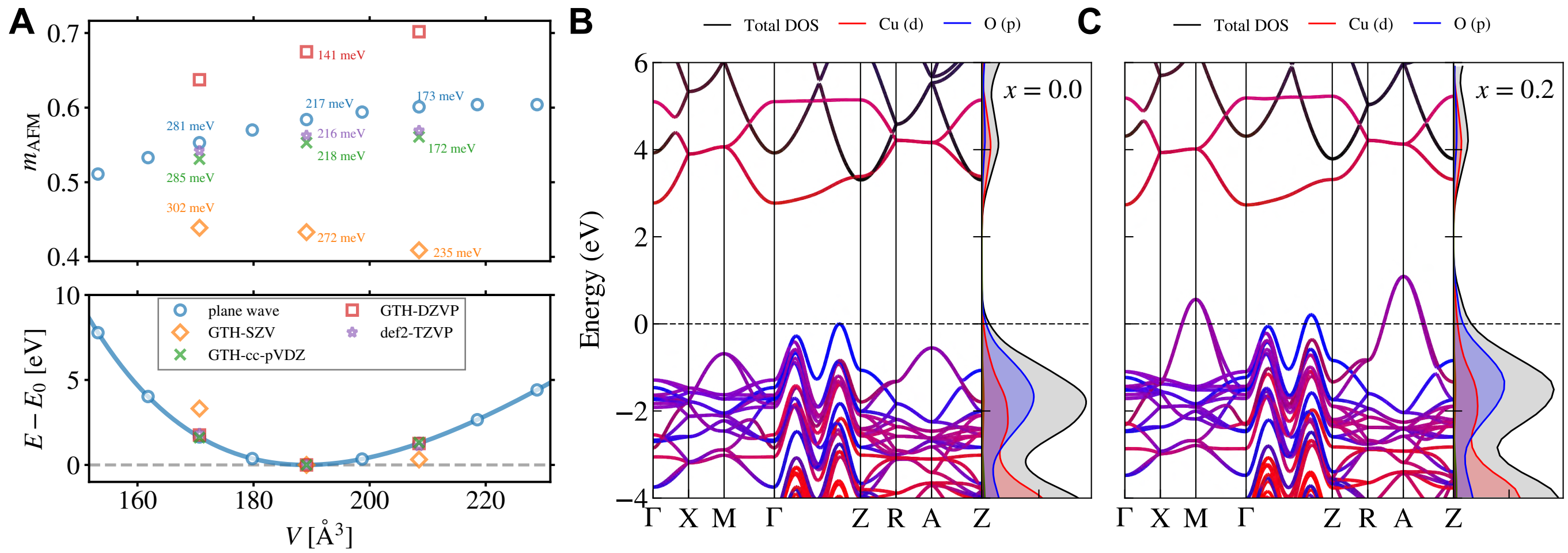

Figure S2: (A) Mean-field basis check of the equation of state, magnetic moment $m_{\mathrm{AFM}}$ and magnetic exchange coupling $J$ of CCO using unrestricted PBE0. Corresponding $J$'s are given by the labels. (B) and (C): Band structure (in the $2 \times 2$ cell Brillouin zone) and projected density of states (DOS) of CCO at doping $x = 0.0$ and 0.2, using PBE0 and the rigid band approximation.

## 2.3 Mean-field settings

The single particle mean-field (SCF) calculations (HF, DFT) were carried out using the crystalline Gaussian atomic orbital basis representation in the PySCF package [91, 92]. The results of these calculations were cross-checked against plane wave basis calculations using the VASP package [93, 94, 95, 96, 97].

For CCO and the Hg-based cuprates, we used the correlation consistent double-$\zeta$ basis GTH-cc-pVDZ defined above, using the GTH pseudopotential for the core electrons [84, 85]. Gaussian density fitting was used to compute the two-electron integrals. We used a specially-optimized Gaussian basis as the density fitting auxiliary basis for Cu, O and Ca; and used the def2-TZVP-RI basis for the auxiliary basis of Hg and Ba ($n_{\mathrm{aux}} \sim 5 n_{\mathrm{AO}}$).

For the plane wave basis calculations, a projector-augmented wave (PAW) [98, 97] representation was used to treat the core electrons and we used a plane wave kinetic energy cutoff of 500 eV.

We sampled the Brillouin zone with a $\Gamma$-centered $\mathbf{k}$ mesh: we used $4 \times 4 \times 2$ for the $2 \times 2$ supercell of the single layer compounds CCO and Hg-1201; $4 \times 4 \times 1$ for the $2 \times 2$ supercell of the double-layer compound Hg-1212.

All mean-field calculations used a Fermi-Dirac smearing of 0.2 eV. All mean-field calculations were converged to an accuracy of better than $10^{-8}$ a.u. per unit cell.

We used the PBE0 [99] hybrid functional for all doping RBA and VCA concentrations. All calculations were spin-unrestricted so that the AFM order could be stabilized at the DFT level. HF calculations were also performed, but we found that these gave very poor descriptions of the doped states.

In general, it is important to note that the basic features of the charge density are established by the choice of mean-field and the choice of doping representation rather than the DMET self-consistency, as the self-consistency is performed here on the anomalous part of the correlation potential.

### 2.4 DMET settings

All DMET routines, including the bath construction, integral transformation, solver interface, and correlation potential fitting, were implemented in the LIBDMET package [62, 61]. To remove core orbitals (which make the bath construction unstable and increases the computational cost) we froze the lowest mean-field bands ($3s3p$ bands for Cu and Ca, $2s$ bands for O, $5s$ bands for Ba); and we also froze the Cu $4f$ and O $3d$ virtuals to further reduce the cost.

We added the correlation potential $u$ to the $CuO_2$ three-band orbitals and *only* fitted the three-band orbital anomalous blocks of the density matrices, i.e., $\langle a_{i\alpha} a_{j\beta} \rangle$ where $ij \in$ three-band orbitals (the self-consistency of the normal magnetic part is not considered in this work). We also enforced $C_{2h}$ symmetry in $u$. The initial guess of $u$ was chosen as a $d$-wave pattern on Cu $3d_{x^2-y^2}$ orbitals with a small amplitude $10^{-3}$. The convergence criterion on the DMET self-consistency was chosen such that the maximal change of an element in $u$ was less than $5 \times 10^{-4}$ a.u..

In the DMET mean-field and correlation fitting, a small smearing of $\beta = 1/k_{\mathrm{B}}T = 1000$ a.u. was added to the lattice.

We used the Newton-Krylov GCCSD methods implemented in MPI4PYSCF [74] as solvers. The CCSD $T$ and $\Lambda$ equations were converged to a residual of less than $10^{-4}$ a.u. The largest embedding problem we treated using the GCCSD solver was of size (376o, 188e), with multiple such size fragments solved simultaneously in the multi-fragment embedding formalism.

For TCCSD, we use an active space of 16 spin orbitals, defined from the CCSD natural orbitals. The active space was solved by the exact diagonalization.

Ab initio DMRG in the generalized spin orbital formalism was implemented in BLOCK2 package [100]. A series of active spaces of 16, 32, 48, 64 orbitals (using CCSD natural orbitals in the last DMET iteration) were solved by DMRG with a maximal bond dimension $M = 3000$.

## 3 Additional data

### 3.1 Ab initio impurity solver benchmarks and order parameters

Fig. S3 shows AFM order and $d$-wave pairing order computed using a $2 \times 2$ plaquette impurity embedded in the 1-band 2D Hubbard model using an exact diagonalization (FCI) solver and the CCSD solver. (Note that this data is intended to test the quality of the CCSD solver, rather than report on the detailed physics of the 2D Hubbard model). As can be seen, the CCSD solver provides an accurate impurity solver for the magnetic order at all dopings. The pairing order from the CCSD solver is slightly less accurate, but still shows quite good agreement with the exact diagonalization solver up to $U = 6$. The main discrepancy from ED at $U = 8$ is the width of the SC dome, which is narrower when using the CCSD solver.

In Fig. S4A we show the Cu $d$-wave order parameter as a function of doping for CCO at ambient pressure using several different solvers. The full orbital space for the CCO impurity consists of 376 orbitals. In the DMRG calculations, we solve the impurity problem from the last iteration of the DMET self-consistency (using the CCSD solver), but reduce the impurity problem orbital size (using coupled cluster natural orbitals) to active spaces of 16-64 orbitals. We see that the order parameter produced by DMRG is in good agreement with that obtained from CCSD. The TCCSD calculations used a 16-orbital active space to fix the active space amplitudes, but otherwise correlated all orbitals, and were performed with full DMET self-consistency. We see that these are also in good agreement with the results from the CCSD solver performed with DMET self-consistency. Taken together with the benchmark data generated on the Hubbard model on SC order with the CCSD solver, as well as multiple studies of the accuracy of the CCSD solver for magnetic order parameters in transition metal oxide materials [101, 25, 102, 62], we expect the trends in this work to be adequately reproduced using the CCSD solver.

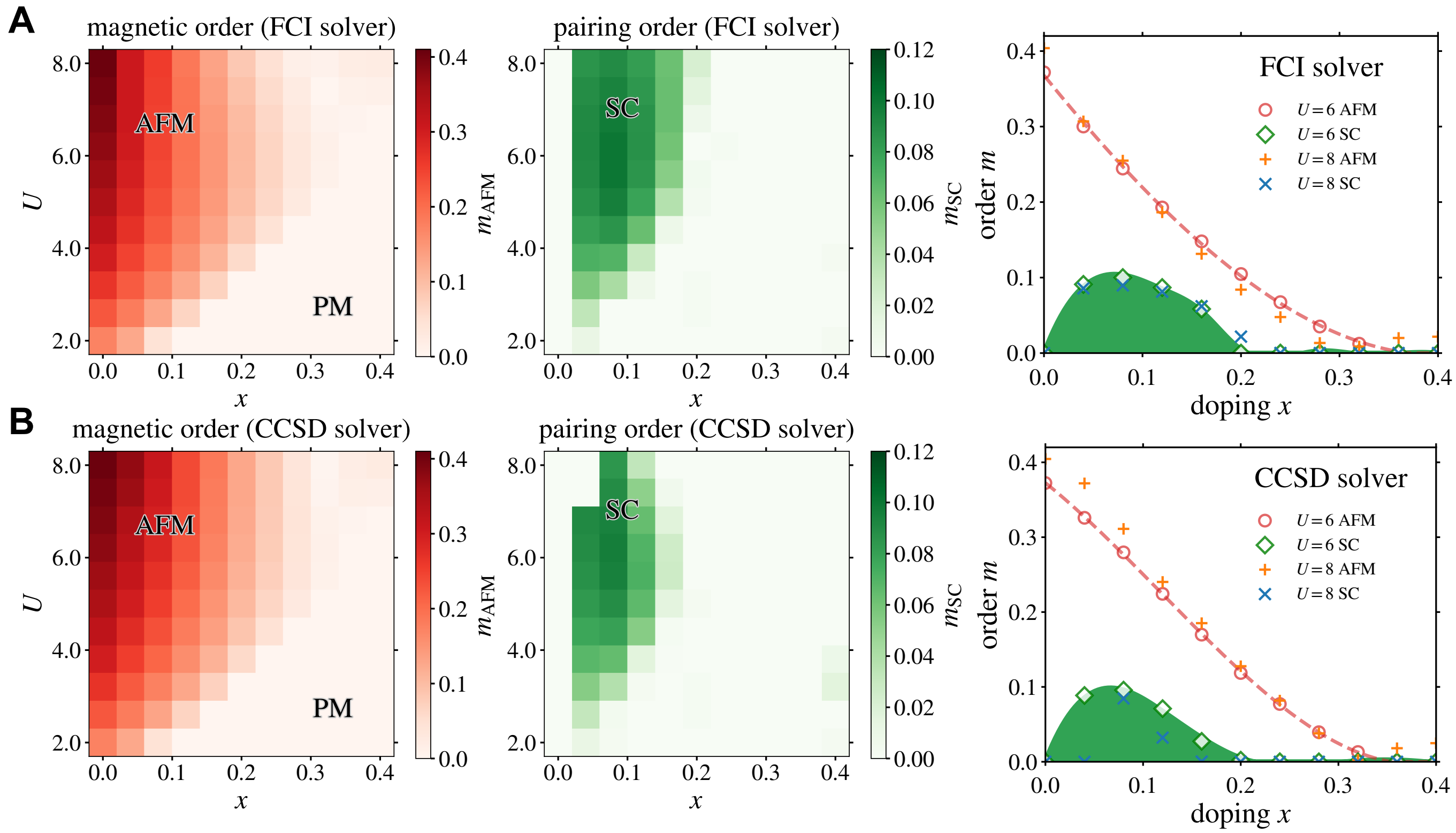

Figure S3: **Benchmark of impurity solver on the phase diagram of the one-band Hubbard model.** A $2 \times 2$ impurity is embedded in $40 \times 40$ square lattice. The figure plots the magnetic (AFM, paramagnetic (PM)) and pairing order ($d$-wave) for different $U$ and doping $x$ from (A) FCI (exact diagonalization) solver and (B) CCSD solver. The $U = 6$ and $U = 8$ orders are also shown in the third column for clarity.

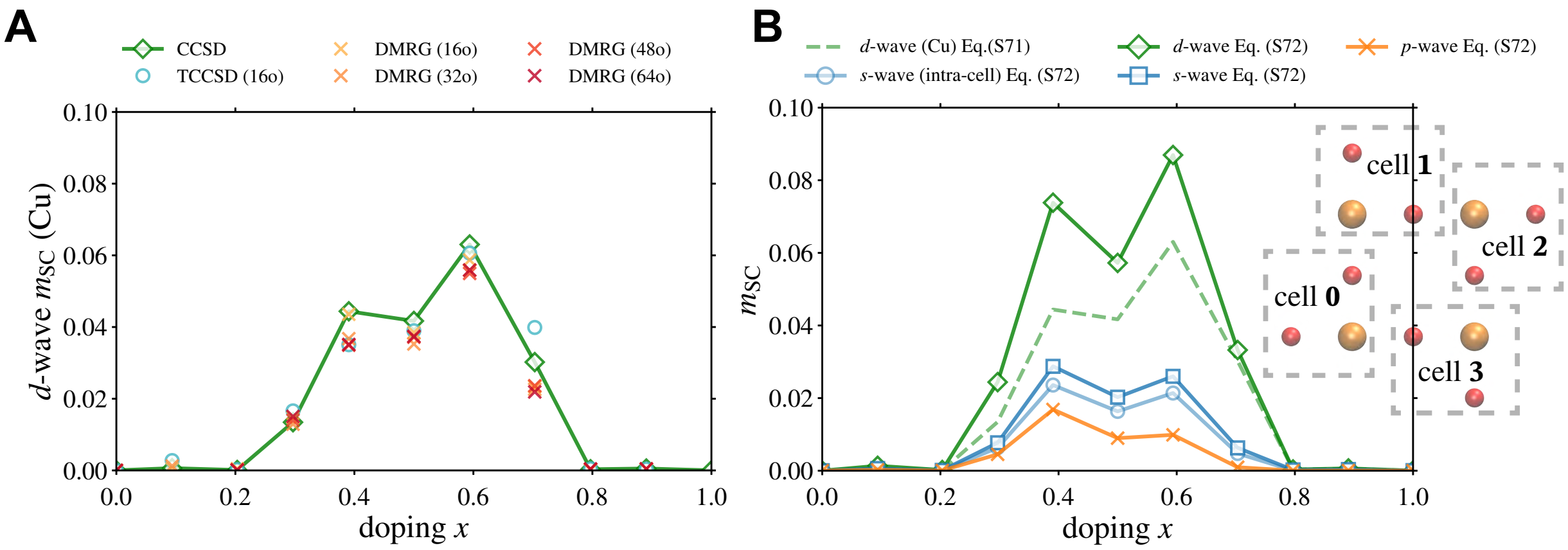

Figure S4: **Benchmark of impurity solver and definitions of order parameter.** (A) Comparison between different solvers (for CCO at normal pressure): CCSD, tailored CCSD with active space of 16 orbitals. DMRG active space calculations of 16 - 64 orbitals. The active spaces are spanned from the CCSD natural orbitals near the Fermi level. (B) Comparison between different definitions of pairing order parameter. Using Eq. (S72), we calculated $s$, $p$, $d$-wave component of order parameters among 4 unit cells in the $2 \times 2$ impurity. The $d$-wave pairing order between 4 Cu's (using Eq. (S71)) is shown by the dashed line.

In Fig. S4B we show additional data giving the decomposition of the total pairing order into intra-cell and inter-cell components in CCO at ambient pressure (using Eq. (S72)). In Fig. S5 we show the orbital-resolved $d$-wave orders (using Eq. (S71)).

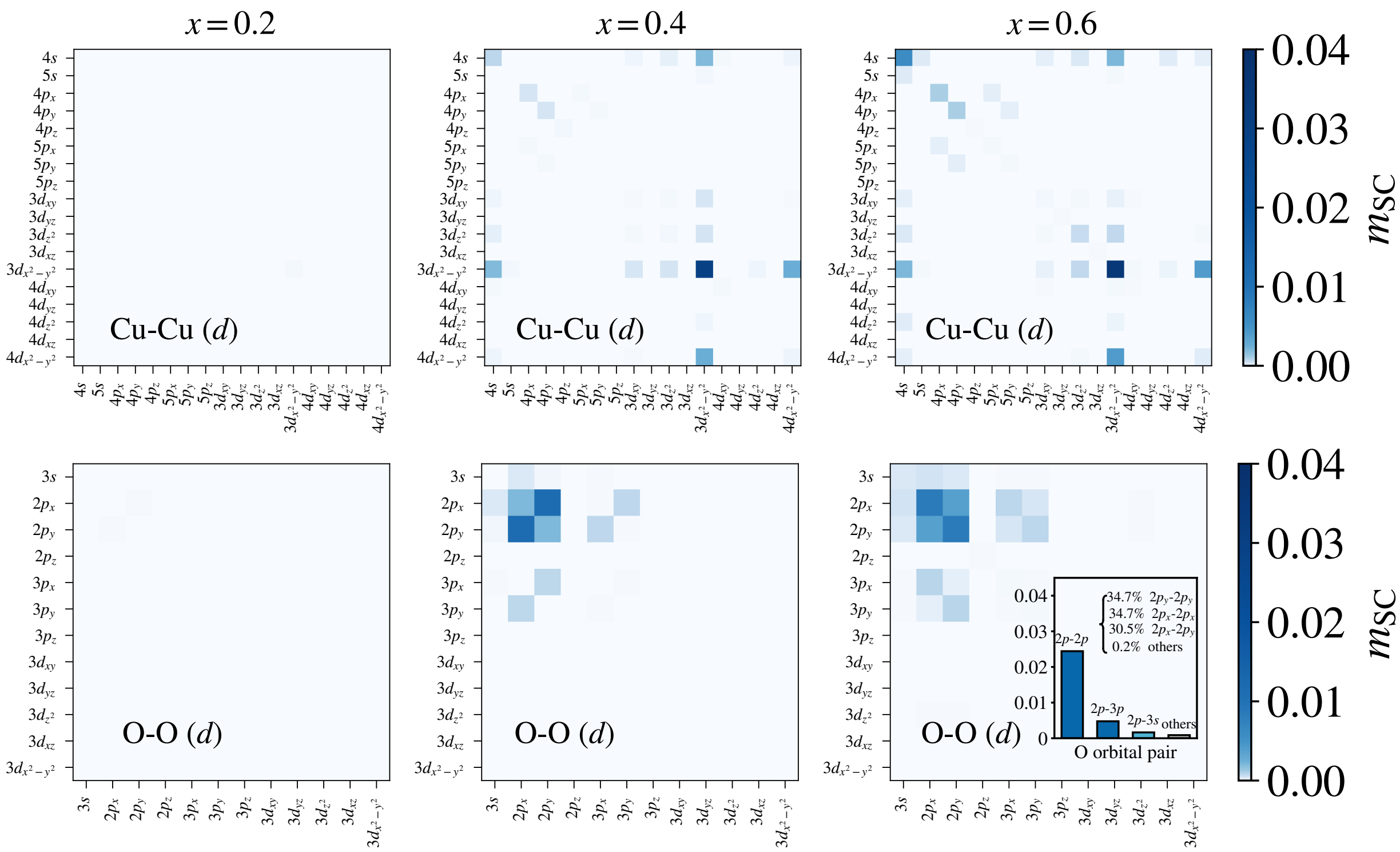

Figure S5: Orbital-resolved $d$-wave order parameter of CCO at ambient pressure for Cu-Cu and O-O orbital pairs ($4f$ orbitals are omitted for clarity). The inset shows the percentage contribution of different kinds of O-O pairing, similar to Fig. 2D in the main text.

### 3.2 Virtual crystal approximation data

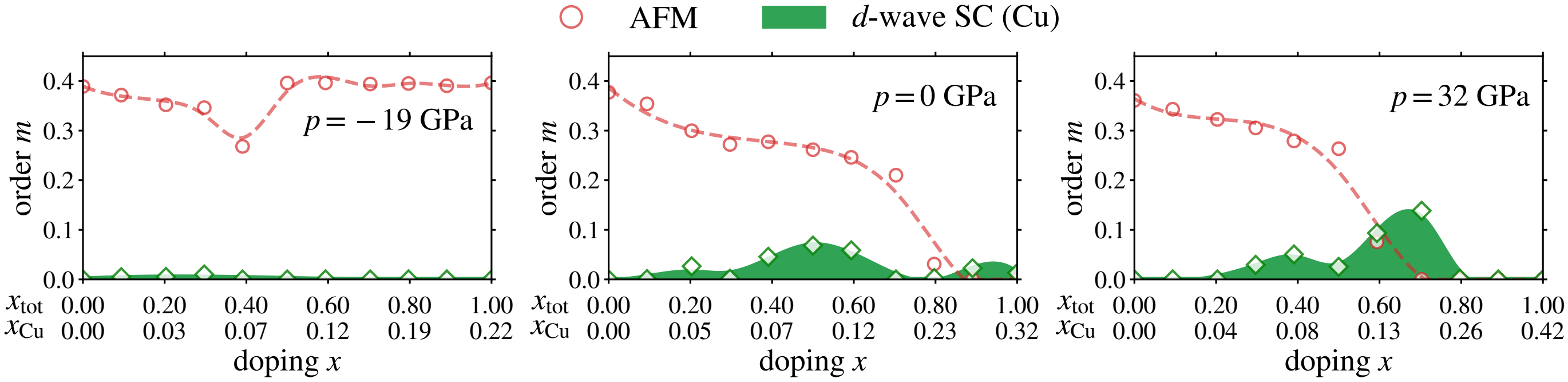

Figure S6: Pressure effect on hole-doped CCO using DMET @ VCA.

In Fig. S6 we show results from the VCA for CCO at different pressures. The general features, such as the decay of the AFM order, the emergence of the SC dome(s), the increas of the SC strength under pressure, are almost the same as with the RBA (Fig. 2 in the main text). This indicates the robustness of the pressure trend. One interesting feature is that in the VCA, the SC order is very weak in the negative pressure case, possibly owing to the fact that fewer holes go to the $CuO_2$ plane as compared to the RBA, leading to a lower effective Cu doping.

### 3.3 Influence of lattice smearing temperature

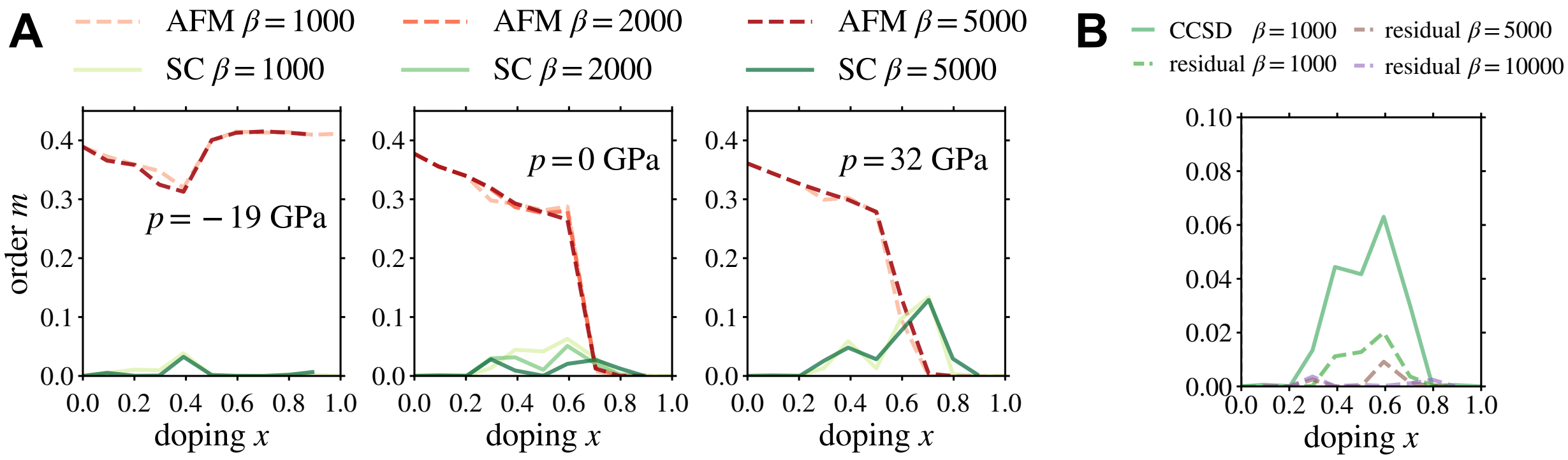

Figure S7: (A) Results of CCO using different lattice temperatures $\beta = \frac{1}{k_B T} = 1000, 2000, 5000$ a.u.. (B) The Hartree-Fock pairing residual from the mismatch between the finite temperature lattice problem ($\beta = 1000, 5000, 10000$) and zero-temperature impurity problem. The CCSD result at $\beta = 1000$ is also shown for comparison.

The finite temperature smearing is used to help the DMET fitting and self-consistency converge: in general, the fitting and convergence become harder at lower temperatures. We see in Fig. S7(A) that there are some numerical differences at ambient pressure due to the finite temperature smearing, however, the basic trend in maximum pairing order is reasonably robust.

Due to the mismatch between the finite smearing temperature of the lattice problem and the 0 K calculation of the impurity problem, the DMET self-consistent cycle produces a non-zero correlation potential to match the (zero-temperature) impurity and (finite-temperature) lattice density matrices, even when using a Hartree-Fock solver in the impurity (when there should be no correlation). The residual correlation potential, and the residual pairing order that it generates, go to zero as the lattice temperature goes to 0. Fig. S7(B) shows the residual pairing going to zero as $\beta \to \infty$, and that it is always significantly smaller than the pairing that is generated when using an actual correlated impurity solver.

There is also an interplay between the lattice smearing temperature and the finite size effect of the DMET lattice. As we have a finite **k**-mesh, capturing the metallic state in the mean-field calculation requires a smearing inversely proportional to the discretization of the band energies implied by the **k**-mesh. We see that at low temperatures, this finite **k**-mesh manifests itself as a dip in the SC order around $x = 0.5$. At high pressure, this dip appears at higher temperatures, due to the larger bandwidth (and thus larger energy discretization for a fixed **k**-mesh).

### 3.4 Comparison between reduced and oxidized structures of Hg-1212

Table S4: Properties of reduced (red) and oxidized (ox) structures of Hg-1212 using DFT (PBE0).

| Structure | Hg-1212 (red) | | Hg-1212 (ox) | |
|---|---|---|---|---|
| doping | $x = 0.0$ | $x = 0.5$ | $x = 0.0$ | $x = 0.5$ |
| $V$ [Å$^3$] | 757.9 | | 752.9 | |
| $d$ Cu-O in-plane [Å] | 1.932 | | 1.928 | |
| $d$ Cu-O apical [Å] | 2.824 | | 2.775 | |
| $m_{\text{AFM}}$ (**k**-$4 \times 4 \times 1$ PySCF) | 0.560 | 0.429 | 0.557 | 0.401 |
| $m_{\text{AFM}}$ (**k**-$4 \times 4 \times 1$ VASP) | 0.592 | 0.350 | 0.591 | 0.327 |
| $m_{\text{AFM}}$ (**k**-$4 \times 4 \times 2$ VASP) | 0.590 | 0.350 | 0.590 | 0.334 |
| $m_{\text{AFM}}$ (**k**-$6 \times 6 \times 4$ VASP) | 0.591 | 0.358 | 0.590 | 0.327 |
| $\Delta n$ (in-plane) | - | -2.576 | - | -2.698 |
| $\Delta n$ (out-of-plane) | - | -1.424 | - | -1.302 |

### 3.5 Coupled-cluster amplitude decomposition data

Table S5: SVD decomposition of $T_2$ amplitudes. Singular values $\Sigma$ and singular vector component (excitation) norms of the leading 5 modes are shown.

| | spin-conserving | | | spin-fluctuating | | |
|---|---|---|---|---|---|---|
| mode | $\Sigma$ | excitation [multiplicity] | norm | $\Sigma$ | excitation [multiplicity] | norm |
| 1 | 0.366 | | | 0.541 | | |
| | | Cu $3d_{x^2-y^2}$ $\alpha \rightarrow$ O $2p_y$ $\alpha$ [×2] | 0.023 | | Cu $3d_{x^2-y^2}$ $\alpha \rightarrow$ Cu $3d_{x^2-y^2}$ $\beta$ [×4] | 0.074 |
| | | Cu $3d_{x^2-y^2}$ $\beta \rightarrow$ O $2p_x$ $\beta$ [×2] | 0.023 | | O $2p_x$ $\alpha \rightarrow$ O $2p_x$ $\beta$ [×2] | 0.020 |
| | | O $2p_y$ $\beta \rightarrow$ Cu $3d_{x^2-y^2}$ $\beta$ [×2] | 0.008 | | O $2p_y$ $\alpha \rightarrow$ O $2p_x$ $\beta$ [×4] | 0.008 |
| | | O $2p_x$ $\alpha \rightarrow$ Cu $3d_{x^2-y^2}$ $\alpha$ [×2] | 0.008 | | O $2p_x$ $\alpha \rightarrow$ O $2p_x$ $\beta$ [×4] | 0.008 |
| | | Cu $3d_{yz}$ $\beta \rightarrow$ Cu $4d_{yz}$ $\beta$ [×2] | 0.007 | | … | … |
| | | Cu $3d_{xz}$ $\alpha \rightarrow$ Cu $4d_{xz}$ $\alpha$ [×2] | 0.007 | | | |
| | | … | … | | | |
| 2 | 0.362 | | | 0.453 | | |
| | | Cu $3d_{x^2-y^2}$ $\alpha \rightarrow$ O $2p_y$ $\alpha$ [×2] | 0.031 | | Cu $3d_{x^2-y^2}$ $\alpha \rightarrow$ Cu $3d_{x^2-y^2}$ $\beta$ [×4] | 0.117 |
| | | O $2p_y$ $\beta \rightarrow$ Cu $3d_{x^2-y^2}$ $\beta$ [×2] | 0.020 | | O $2p_x$ $\alpha \rightarrow$ O $2p_x$ $\beta$ [×4] | 0.016 |
| | | Cu $3d_{x^2-y^2}$ $\beta \rightarrow$ O $2p_x$ $\beta$ [× 2] | 0.015 | | O $2p_y$ $\alpha \rightarrow$ O $2p_x$ $\beta$ [×4] | 0.007 |
| | | O $2p_y$ $\alpha \rightarrow$ Cu $3d_{x^2-y^2}$ $\alpha$ [× 2] | 0.013 | | O $2p_x$ $\beta \rightarrow$ O $2p_y$ $\beta$ [×2] | 0.006 |
| | | Cu $3d_{yz}$ $\beta \rightarrow$ Cu $4d_{yz}$ $\beta$ [× 2] | 0.011 | | … | … |
| | | Cu $3d_{xz}$ $\beta \rightarrow$ Cu $4d_{xz}$ $\beta$ [× 2] | 0.011 | | | |
| | | … | … | | | |
| 3 | 0.362 | | | 0.453 | | |
| | | Cu $3d_{x^2-y^2}$ $\beta \rightarrow$ O $2p_x$ $\beta$ [× 2] | 0.031 | | Cu $3d_{x^2-y^2}$ $\alpha \rightarrow$ Cu $3d_{x^2-y^2}$ $\beta$ [×4] | 0.073 |
| | | O $2p_x$ $\alpha \rightarrow$ Cu $3d_{x^2-y^2}$ $\alpha$ [× 2] | 0.020 | | O $2p_y$ $\alpha \rightarrow$ O $2p_x$ $\beta$ [×4] | 0.021 |
| | | Cu $3d_{x^2-y^2}$ $\alpha \rightarrow$ O $2p_y$ $\alpha$ [× 2] | 0.015 | | O $2p_x$ $\alpha \rightarrow$ O $2p_x$ $\beta$ [×4] | 0.013 |
| | | O $2p_x$ $\beta \rightarrow$ Cu $3d_{x^2-y^2}$ $\beta$ [× 2] | 0.013 | | O $2p_y$ $\beta \rightarrow$ O $2p_x$ $\beta$ [×2] | 0.010 |
| | | Cu $3d_{xz}$ $\alpha \rightarrow$ Cu $4d_{xz}$ $\alpha$ [× 2] | 0.011 | | … | … |
| | | Cu $3d_{yz}$ $\alpha \rightarrow$ Cu $4d_{yz}$ $\alpha$ [× 2] | 0.011 | | | |
| | | … | … | | | |
| 4 | 0.361 | | | 0.344 | | |
| | | Cu $3d_{x^2-y^2}$ $\beta \rightarrow$ O $2p_x$ $\beta$ [× 2] | 0.020 | | Cu $3d_{x^2-y^2}$ $\alpha \rightarrow$ Cu $3d_{x^2-y^2}$ $\beta$ [×4] | 0.074 |
| | | Cu $3d_{x^2-y^2}$ $\alpha \rightarrow$ O $2p_y$ $\alpha$ [× 2] | 0.020 | | O $2p_x$ $\alpha \rightarrow$ O $2p_x$ $\beta$ [×2] | 0.020 |
| | | O $2p_y$ $\alpha \rightarrow$ Cu $3d_{x^2-y^2}$ $\alpha$ [× 2] | 0.014 | | O $2p_y$ $\alpha \rightarrow$ O $2p_x$ $\beta$ [×4] | 0.008 |
| | | O $2p_x$ $\beta \rightarrow$ Cu $3d_{x^2-y^2}$ $\beta$ [× 2] | 0.014 | | O $2p_x$ $\alpha \rightarrow$ O $2p_x$ $\beta$ [×4] | 0.008 |
| | | O $2p_y$ $\beta \rightarrow$ Cu $3d_{x^2-y^2}$ $\beta$ [× 2] | 0.014 | | … | … |
| | | O $2p_x$ $\alpha \rightarrow$ Cu $3d_{x^2-y^2}$ $\alpha$ [× 2] | 0.014 | | | |
| | | … | … | | | |
| 5 | 0.270 | | | 0.227 | | |
| | | O $2p_y$ $\alpha \rightarrow$ Cu $3d_{x^2-y^2}$ $\alpha$ [× 2] | 0.009 | | O $2p_x$ $\alpha \rightarrow$ O $2p_x$ $\beta$ [×4] | 0.020 |
| | | O $2p_x$ $\beta \rightarrow$ Cu $3d_{x^2-y^2}$ $\beta$ [× 2] | 0.009 | | O $2p_y$ $\alpha \rightarrow$ O $2p_x$ $\beta$ [×4] | 0.012 |
| | | O $2p_y$ $\beta \rightarrow$ Cu $3d_{x^2-y^2}$ $\beta$ [× 2] | 0.009 | | O $2p_x$ $\alpha \rightarrow$ Cu 4px $\beta$ [×4] | 0.011 |
| | | O $2p_x$ $\alpha \rightarrow$ Cu $3d_{x^2-y^2}$ $\alpha$ [× 2] | 0.009 | | O $2p_y$ $\alpha \rightarrow$ O $2p_y$ $\beta$ [×2] | 0.010 |
| | | O $2p_y$ $\beta \rightarrow$ O $2p_y$ $\beta$ [× 2] | 0.006 | | … | … |
| | | O $2p_x$ $\alpha \rightarrow$ O $2p_x$ $\alpha$ [× 2] | 0.006 | | | |
| | | … | … | | | |

### 3.6 Effect of different fluctuations (diagrams) in coupled-cluster theory on the superconductivity

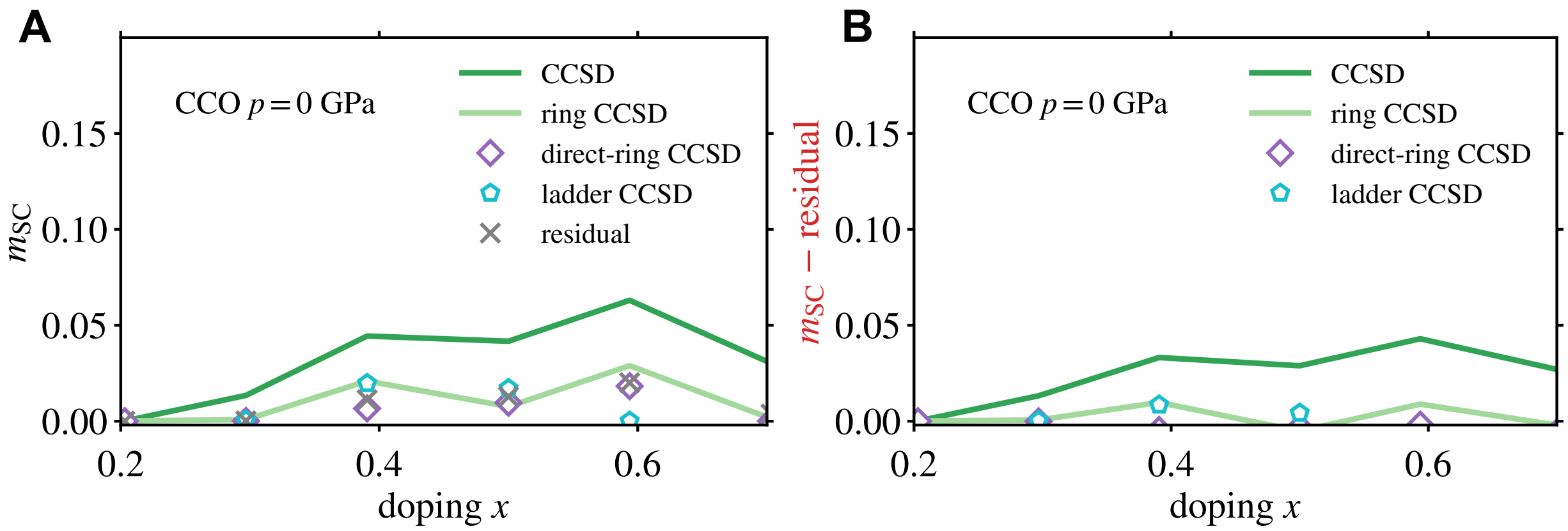

Figure S8: Superconducting pairing orders of CCO (at ambient pressure) from different CC variants: full CCSD, ring CCSD, direct-ring CCSD, and ladder CCSD. A residual SC order associated with the lattice smearing temperature $\beta = 1000$ (see Fig. S7 for details) has been subtracted in (B) to show the pure correlation effect in the mechanism. The residual order results from the $\beta$ mismatch between the lattice and impurity and goes away as $\beta \to \infty$ (see Fig. S7).

### 3.7 Estimation of superconducting gap

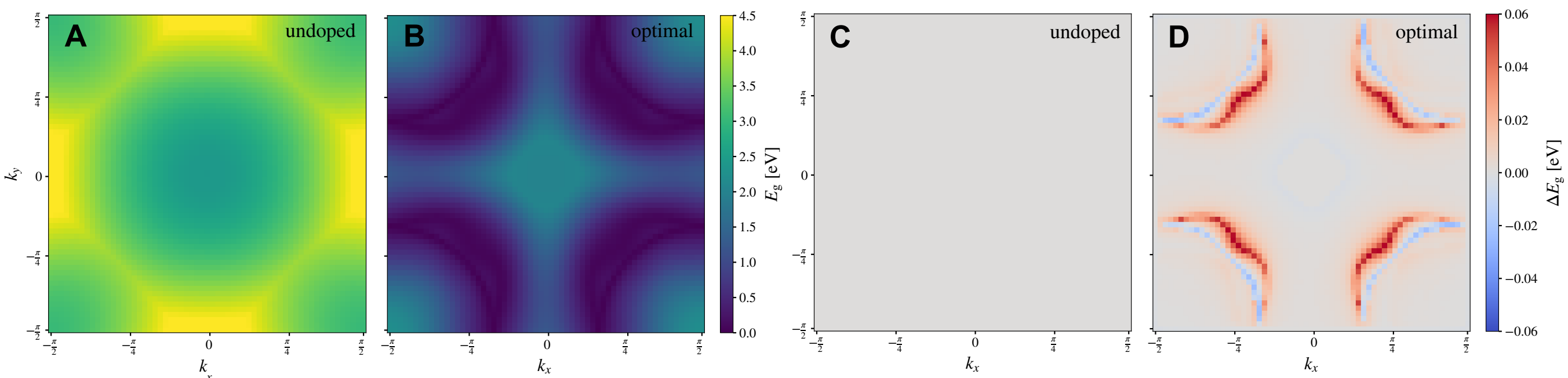

Figure S9: Quasiparticle gap $E_{\mathrm{g}}(\mathbf{k}) = \varepsilon_{\mathbf{k}}^{\mathrm{LUMO}} - \varepsilon_{\mathbf{k}}^{\mathrm{HOMO}}$ from the DMET lattice mean-field of CCO. Gaps $E_{\mathrm{g}}$ of (A) undoped and (B) optimally doped states. (C) and (D): Change of gap compared to the non-superconducting bands at zero and optimal doping. Note that the Brillouin zone is defined with respect to the $2 \times 2$ pinwheel supercell used in the ab initio SC calculations.

To estimate the superconducting gap (which is an experimentally accessible quantity), we plot the HOMO-LUMO band gap in the $2 \times 2$ pinwheel supercell Brillouin zone in Fig. S9. The change of the gap with respect to the non-superconducting state is also shown. We use Wannier interpolation to obtain a finely sampled Brillouin zone.

It is important to sound a note of caution: the HOMO-LUMO gap is not a rigorous approximation to the exact gap of the problem; using it is analogous to using the DFT bandgap to provide information on the true gap, and just as with the DFT bandgap, the DMET HOMO-LUMO gap can exhibit artifacts. With this caveat, however, we also recognize that the DFT bandgap can be quite useful, and we proceed with the HOMO-LUMO DMET gap in this spirit.

In the undoped state, the system is gapped in the whole Brillouin zone and the minimal gap is about 1.7 eV, which is comparable to experimentally measured gaps of related cuprates of about 1.5 - 2 eV [103, 104]. After doping, the system becomes metallic and a pairing gap is opened after adding the correlation potential. Since the energy spectrum is symmetric in the Nambu representation, the largest SC energy gap approximately corresponds to $2\Delta_0$ where $\Delta_0$ is the 0 K gap parameter in the standard 1-band BCS theory.

In the 1-band BCS $d$-wave model, $\Delta_0$ has a direct relation to $T_c$ [68],

$$E_g \sim 2\Delta_0 \approx 4.28 T_c \tag{S82}$$

The largest SC gap $E_g$ in Fig. S9 is about 0.065 eV in the converged DMET lattice solution, which corresponds to a $T_c$ of 180 K from the above relation. Given the various approximations in our methodology and the approximate nature of this $T_c$ relation, we consider this to be in good (perhaps fortuitously good) agreement with the typical cuprate $T_c$'s.

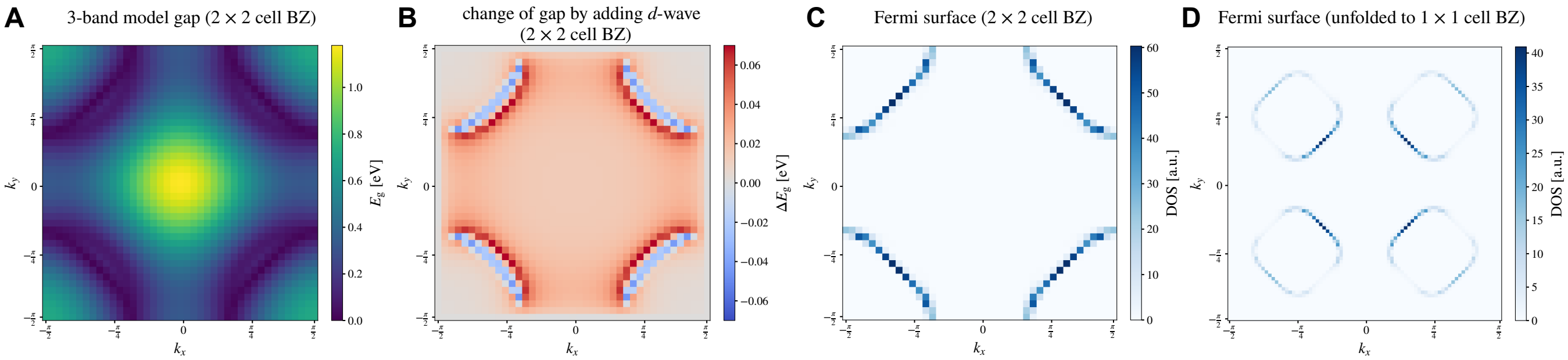

Figure S10: Quasiparticle gap and Fermi surface of a doped 3-band Hubbard model (Hanke parameters [105], $x = 0.3$) with a $d$-wave potential. (A) The quasiparticle gap (B) Change of gap after adding the $d$-wave pairing potential. (C) Fermi surface (defined as the DOS at $\mu = 0$) in the $2 \times 2$ cell BZ. (D) Fermi surface unfolded to the $1 \times 1$ cell BZ.

In the above analysis, the **k**-space distribution of the gap function is based on the $2 \times 2$ cell folded Brillouin zone. It can be easier to understand the physics by unfolding the Brillouin zone to the $1 \times 1$ unit cell [106]. We note that general features of this Fermi surface (such as Fermi pockets), in the presence of AFM order, have been discussed in Ref. [107]. This can be done more easily in the 3-band Hubbard model, to which we have added a pure $d$-wave potential ($\Delta = \pm 0.1$ eV on all off-diagonal Cu-Cu matrix elements. The sign is $+(-)$ if the bond is horizontal(vertical)). From Fig. S10(A) and (B), we see the quasiparticle gap and the change of gap are very similar to the ab initio cases (Fig. S9 (B) and (D)), verifying the $d$-wave nature of the ab initio solution. One can plot the Fermi surfaces by evaluating the density of states at $\mu = 0.0$ and at different **k** points

$$\mathrm{DOS}(\mathbf{k}, \mu = 0) = \frac{1}{\sigma\sqrt{2\pi}} \sum_m \exp\left(-\frac{1}{2}\frac{\varepsilon_{\mathbf{k}m}^2}{\sigma^2}\right), \tag{S83}$$

where $\sigma = 0.1$ eV is a smearing parameter. Note that because the $2 \times 2$ cell unfolding is not perfect (i.e. the density breaks the translational symmetry of the primitive cell) there can be some artifacts in the unfolded density of states, but we can regard a large value of the density of states as corresponding to a vanishing gap at that point.

Finally, we also computed the superconducting lattice gap as a function of doping $x$ for all compounds we considered (CCO at different pressures, Hg-1201, Hg-1212 (ox) and Hg-1212 (red)), see Fig. 2 and Fig. 3 in the main text). The gap values match well with the SC orders from the anomalous density matrix. In particular, the trends of $E_g$ among the different pressures and number of layers parallel the trends from the SC order parameters, providing support for the arguments in the text.

### 3.8 DFT functional dependence and effects of charge self-consistency

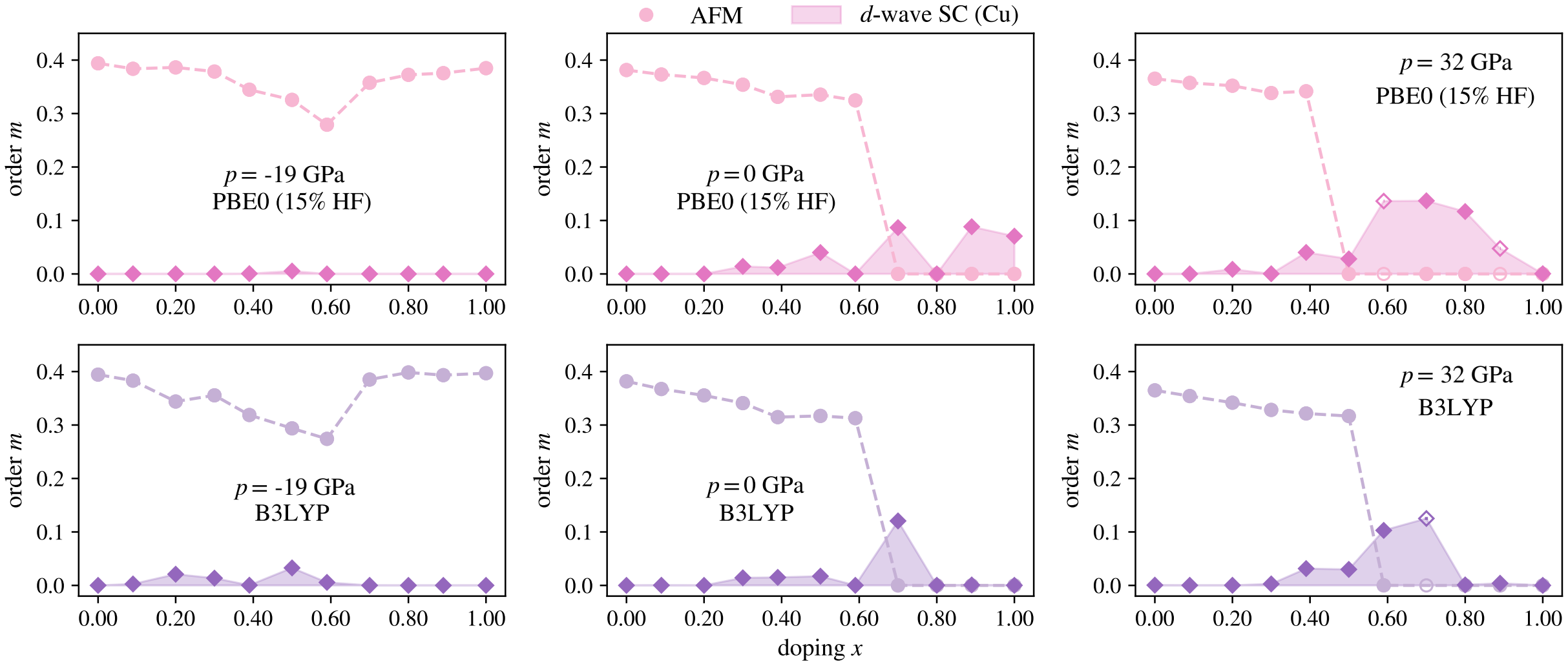

Figure S11: AFM and SC order of CCO with different starting density functionals: PBE0 with 15% HF exchange and B3LYP. The unfilled markers label points where the DMET iterations were slow to converge; the reported orders were then obtained by averaging from the last few iterations (this gives an error bar for the SC order, but is not very visible as it is smaller than 0.01).

As we have discussed in the main text, our current DMET scheme retains some dependence on the initial charge density, because we have not implemented full charge self-consistency. The DMET iterative scheme is not fully charge self-consistent in two ways (i) the non-pairing part of the correlation potential is not fitted in the DMET procedure, and (ii) more importantly, for computational cost reasons, the charge density of the lattice mean-field calculation is not updated, as this would necessitate lattice mean-field (essentially exact exchange) calculations every DMET iteration. Usually, more DMET iterations are also required when CSC is applied [62]. (We do not consider it advisable to perform (i) without (ii), because (i) can lead to a noticeable difference between the impurity charge density and that of the cells that are used to construct the lattice Fock operator).

While a fully charge self-consistent DMET iteration would entirely remove the dependence of DMET on the initial density, and is something we plan for the future, for the purpose of understanding our approximations we now consider the variation of our predictions as we change the initial density.

In the main text we showed results from the standard PBE0 functional. In Fig. S11 we show results from two other hybrid functionals, PBE plus a different fraction (15%) of Hartree Fock exchange, and B3LYP [108, 109], for CCO at different pressures, with RBA. We see that the different functionals yield significant quantitative differences in their predictions. On the other hand, the overall trend of increasing SC order with pressure remains. Consequently, we believe the broad features of the trends we observe are maintained across a range of "reasonable" initial densities.

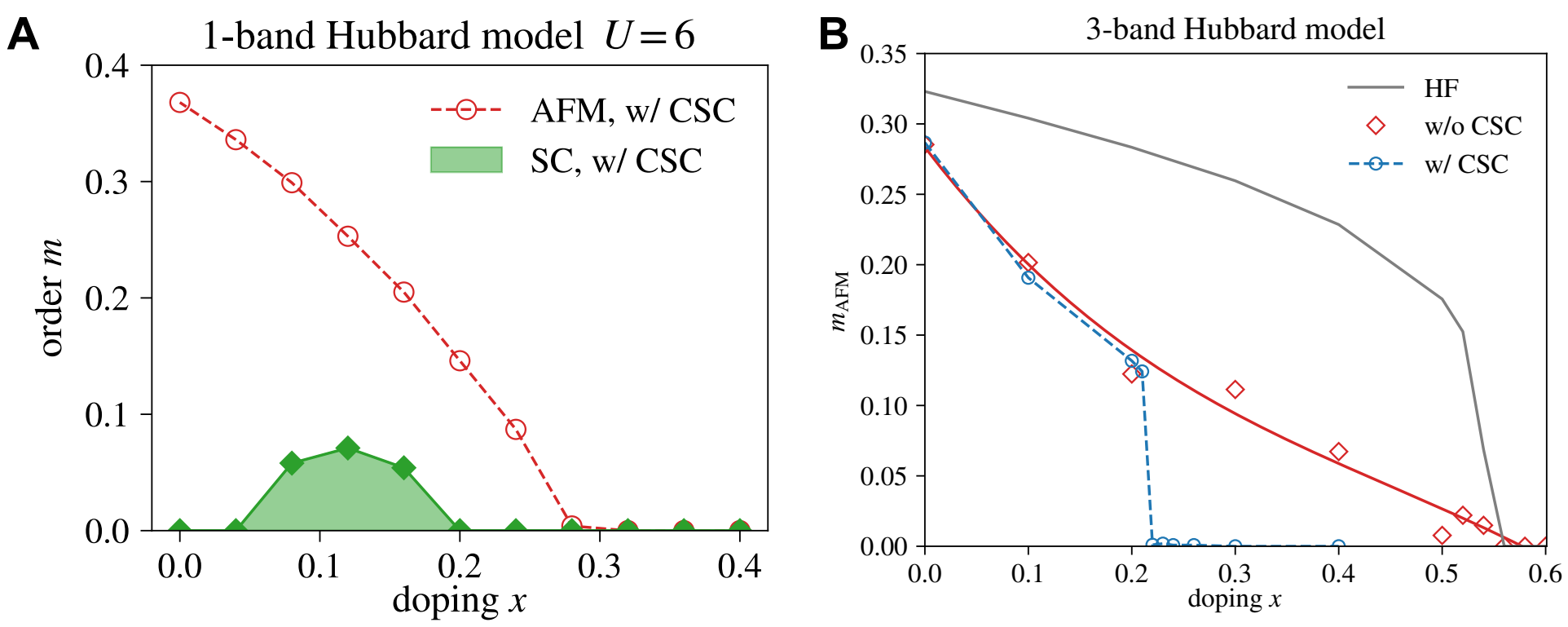

Figure S12: Effects of charge self-consistency (CSC) on (A) 1-band ($U = 6$) and (B) 3-band Hubbard model (using Hanke parameters [105]). The impurity solver is FCI for 1-band model and CCSD for 3-band model. In 3-band model, we find a coexistence of phases between $x = 0.2$ and $x = 0.3$ with CSC and we only show the order from the lowest energy phase. HF denotes the behaviour of the order using Hartree-Fock theory. The solid red curve is cubic-spline interpolated.

One potential consequence of the lack of charge self-consistency discussed in the main text is the large doping needed to remove the magnetic order, which appears closely related to the large doping before we see the onset of SC.

To understand this behavior, we can study the effect of CSC in model problems. In Fig. S12 we show the effects of charge self-consistency (CSC) on 1-band and 3-band Hubbard models.

In the 1-band model, if we perform full charge self-consistency we find that the magnetic moment vanishes below 30% doping. This is not too different from the situation without CSC, where the magnetic moment vanishes at around 35% doping as shown in Fig.S3.

However, in the 3-band model, we find that with CSC, the magnetic moment vanishes at about 22% doping, but without CSC, it decays much more slowly and persists until 56% doping. The slow decay of magnetism without CSC is similar to what is seen in the ab initio results in this work. This suggests that the unphysically large dopings required to quench the magnetism result from the lack of CSC.

The difference between the role of CSC in the 3-band model and the 1-band model gives insight into its effect. Namely, in the 3-band model, CSC allows for a redistribution of charge between copper and oxygen. The spin-polarized bands partially spatially overlap on the oxygen atoms. Thus when they have strong oxygen character near the Fermi surface, holes going into the spin polarized bands (with opposite spin) occupy similar spatial regions, leading to no change in spin polarization. As CSC changes the oxygen content/covalency of the (spin-polarized) bands near the Fermi surface, this changes the amount of uncompensated spin density, and thus the magnetic moment.

# 4 Limitations of the current work

As discussed in the main text, the ab initio methodology developed in the current work reproduces the pressure effect and layer effect in a variety of cuprate materials and structures. Despite this agreement, it is important to note the omissions in the current treatment and areas where the methodology needs to be further improved in the future. These are summarized below.

The main limitations in the basic physics are:

1. Lack of long-range fluctuations. Because we are using a quantum impurity treatment, correlations are limited to being (mainly) across length-scales comparable to the impurity. (In DMET, the correlated fluctuations with the bath provide a limited treatment of longer-range fluctuations outside of the impurity). This rules out pairing mechanisms that rely on long-range or even critical fluctuations. In principle, the pairing order we compute should be carefully converged with respect to the impurity size. This is possible in some cases for models [110] but will clearly be very challenging in a fully ab initio treatment.

2. Lack of explicit doping. As discussed in various places, the connection between chemical doping composition, and effective doping of the plane, is non-trivial and may not be adequately reproduced by the simple RBA and VCA treatments in this work. In addition, dopant atoms introduce disorder which may affect the ground-state order.

3. Lack of phonon degrees of freedom. Phonons are clearly of some importance in these materials, due to strong spin-phonon coupling.

4. Lack of temperature and spectra. Although the current formulation is a ground-state formulation, it is obviously important to include the effects of temperature, as well as to study dynamical effects.

There are also some technical limitations, which appear primarily due to the need to limit computational cost and to ensure numerical stability:

1. Dependence on the initial mean-field. Although the DMET impurity problem does not suffer from double-counting, the DMET self-consistency only modifies a small block of the lattice mean-field Hamiltonian, and elements outside of this block are thus tied to the original mean-field solution, here taken from PBE0. This may not be a good starting point. Indeed, if the charge density from PBE0 is poor we expect poor behavior. As discussed above, the poor charge density of hybrid density functionals under doping is responsible for many of the unusual features that we see in this work, such as persistence of magnetism to large dopings. The solution requires implementing full charge self-consistency in the DMET iterations. This may be the most important technical development to improve this work.

2. The convergence of DMET self-consistency and solvers. The convergence of the DMET self-consistent iteration (and to a lesser extent the many-body solver) is challenging in the doped regime where there are small gaps. The use of finite-temperature smearing here introduces a (small) inconsistency between the lattice treatment (at finite temperature) and the quantum impurity (where the many-body solver works at zero temperature).

3. Treatment of the chemical potential in the solver. Currently, the physical number of electrons is adjusted at the Hartree-Fock level in the quantum impurity for cost reasons, rather than at the correlated level.

4. Non-exactness of the approximate solvers. Although CCSD appears to be a good compromise between accuracy and speed, we may need more powerful solvers to study more exotic phases.

5. Basis set convergence. Although adequate for obtaining the trends in order parameters, the polarized double-zeta basis sets used are still small by the standards of chemical accuracy, where triple-zeta or larger basis sets are desirable.

6. Empirical nature of the gap. While the DMET lattice SC gap provides very suggestive trends for the true SC gap in this work, a more rigorous estimation of the pairing gap is highly desirable (e.g., from computing the Green's function or by removing/adding particles).